\documentclass[a4paper,12pt]{article}
\pdfoutput=1 
\usepackage{jheppub} 
\usepackage[T1]{fontenc} 
\usepackage{amsfonts,amssymb,amsmath}
\usepackage{graphicx,color,float}

\title{
Gravitational corrections to Higgs potentials.} 
\author{Marios Bounakis}
\emailAdd{M.Bounakis2@newcastle.ac.uk}
\author{Ian G. Moss}
\emailAdd{ian.moss@ncl.ac.uk}
\affiliation{School of Mathematics, Statistics and Physics, Newcastle University, 
Newcastle Upon Tyne, NE1 7RU, UK}
\date{\today}

\abstract{
Understanding the Higgs potential at large field values corresponding to
scales in the range above $10^{10}{\rm GeV}$ is important for questions
of vacuum stability, particularly in the early universe where survival of
the Higgs vacuum can be an issue. In this paper we show that the
Higgs potential can be derived in away which is independent of the
choice of conformal frame for the spacetime metric. Questions about
vacuum stability can therefore be answered unambiguously. We show that
frame independence leads to new relations between the beta functions
of the theory and we give improved limits on the allowed values of the
Higgs curvature coupling for stability.
}

\begin{document} 
\maketitle
\section{introduction}

Extrapolation of the Standard Model of particle physics to high energies leads to
the remarkable conclusion that our vacuum is only a long-lived metastable
state, in which the Higgs field sits at a local minimum of the Higgs potential
surrounded by a potential barrier of width somewhere in the range $10^{10}-10^{14}{\rm GeV}$
\cite{Degrassi:2012ry,Buttazzo:2013uya,Blum:2015rpa}. 
This raises an interesting question about initial conditions, because, if the Standard Model is 
correct at these energies, then somehow the Higgs field had to 
evolve into the metastable vacuum state during the early stages of the universe
\cite{Espinosa:2007qp}.

The Higgs potential barrier depends strongly on the effective Higgs mass at high energies,
and it is quite possible that gravitational corrections may be important.
In the relevant energy range, there is no reason to abandon General Relativity as 
`an effective field theory' description of gravity \cite{Burgess:2003jk}. There are two contributions 
to the effective Higgs mass, the ordinary one and the mass due to the coupling 
$\xi R{\cal H}^\dagger{\cal H}$, between the Higgs field ${\cal H}$ and 
the curvature $R$. We will assume that inflation is driven by an inflaton field, not the
Higgs field, which is assumed weakly
interacting and makes no contribution to the Higgs potential. The curvature coupling increases the height of the 
potential barrier around the metastable minimum if $\xi R$ is positive, and has the opposite effect 
when $\xi R$ is negative, making Higgs stability sensitive to the value of $\xi$. 

Placing the Higgs decay into a cosmological context introduces an ambiguity in how we define
the spacetime geometry. In particular, we can perform a conformal re-scaling of the metric which 
removes the curvature-coupling term, transforming the theory from the original 
Jordan frame metric to the Einstein frame metric. It has been noted that quantum calculations can sometimes
lead to different results when done in the Jordan or the Einstein frame \cite{George:2013iia,Steinwachs:2013tr}. 
This is a puzzle, because we want to avoid a situation in which the Higgs field is unstable in the Jordan frame and 
stable in the Einstein frame. The contradiction would be best resolved by having an approach to 
quantisation which is consistent  whatever the choice of spacetime metric
\cite{Kamenshchik:2014waa,George:2015nza,Moss:2014nya,Moss:2015gua}. 
We shall show that there is a covariant quantisation 
scheme which gives consistent results on Higgs instability. 

The basic tool we use is an effective action which is covariant under field 
transformations
\cite{Vilkovisky:1984,DeWittdynamical,barvinsky1985generalized,ParkerTomsbook}. 
This is a stronger requirement than General Covariance, 
or covariance under spacetime coordinate transformations.  The idea of a field-space covariant 
quantum field theory (hereafter called covariant)  is illustrated by the diagram 
in Eq (\ref{diag}). Quantisation followed by a field redefinition should give the same result as starting
from a field redefinition and then quantising, i.e. the diagram should commute.
\begin{equation}
\begin{matrix}
\varphi&\to&\varphi'\\
\downarrow&&\downarrow\\
\Gamma[\varphi]&\to&\Gamma[\varphi']\\
\end{matrix}\label{diag}
\end{equation}
Demanding covariance of the effective action guarantees covariance of the effective field 
equations. Without covariance, there is a different quantum field theory for each
choice of field variables. Imposing covariance has another virtue. Terms are added
to the classical Lagrangian to fix the gauge freedom. Solutions to the usual effective
field equations depend on the choice of these gauge-fixing terms. However, in the covariant
approach, the solutions to the effective field equations are independent
of the gauge-fixing terms.

Covariant approaches are widely used to quantise non-linear sigma models 
\cite{AlvarezGaume198185,Howe:1986vm}, but they
are very rarely used for gauge theories. One reason they are not widely used is that
the gauge-fixing dependence of the usual effective action is not considered
problematic, since the dependence goes away `on-shell' , i.e. the action
takes the same value at solutions to the effective 
field equations \cite{Nielsen1975,Fukuda1976,Kobes:1991,Contreras:1996nx}.
Furthermore, it is easy to show that the Jordan and Einstein frame Higgs theories
have equivalent perturbative expansions when the background fields are on-shell
and the Higgs field is small \cite{Markkanen:2017dlc}.

Using a covariant approach retains the gauge-fixing and frame independence {\em off-shell},
for any range of Higgs field. On the other hand, covariant approaches are not totally
unambiguous, because there are two versions of the 
covariant effective action: $\Gamma[\varphi^*,\varphi]$ which generates the 1PI 
diagrams but depends on an extra field $\varphi^*$ \cite{Burgess:1987zi}, 
and the DeWitt effective action $\Gamma[\varphi]$ which does not generate 
1PI diagrams \cite{DeWitt2}. Fortunately, both generate the same effective field equations, 
and they agree on-shell. They are therefore equivalent for questions of vacuum stability,
so we will use the simpler DeWitt effective action.

Higgs vacuum decay is a situation where the effective action
and the classical action lead to very different qualitative behaviour
\cite{Krive:1977,Polizer:1979,Cabibbo:1979ay,Sher:1988mj}. Another example
is Coleman-Weinberg theory of massless electrodynamics, where quantum
corrections to the effective action lead to symmetry breaking. In these
situations, we use solutions to the renormalisation group corrected
field equations with running coupling constants to determine the vacuum state or 
to calculate tunnelling amplitudes \cite{coleman}. Note that we use `on shell' to refer to fields
which satisfy the effective field equations rather than the classical field equations.
We will investigate whether covariant and non-covariant approaches to the
effective action give different physical results by doing specific calculations of the
running couplings in the Higgs effective potential.

The renormalisation group corrected potential used here is constructed as follows. 
The DeWitt effective action for the modulus of the Higgs field $\phi$ is written as a functional 
$\Gamma(g_i,\phi,g_{\mu\nu},\mu_R)$, where $g_i$ are running
couplings depending on $\mu_R$, the renormalisation scale. At one loop order,
the explicit dependence on renormalisation scale has contributions from
all types of field in the standard model. These contributions are determined by
perturbation theory and depend on a set of second order differential operators 
$\Delta^n(\phi)$. Following Coleman and Weinberg \cite{coleman}, the $\beta$ functions can be
obtained by comparing coefficients in the renormalisation group equation
for the Lagrangian,
\begin{equation}
\sum_i\beta_i{\partial {\cal L}\over\partial g_i}
-\gamma_\phi\phi{\partial {\cal L}\over\partial\phi}-
\gamma_g g_{\mu\nu}{\partial {\cal L}\over\partial g_{\mu\nu}}
={1\over 16\pi^2}\sum_n (\pm) b_2(\Delta^n)\label{scaling},
\end{equation}
where the sign is positive for bosons and negative for fermions and ghosts.
Renormalisation of the fields is responsible for the anomalous dimensions 
$\gamma_\phi$ and $\gamma_g$ (where we are using the sign conventions of \cite{Sher:1988mj}).
The functions $b_2$ are polynomial combinations of coefficients in the operators $\Delta^n$. 
General expressions for $b_2$ are known for many types of operators on arbitrary spacetime backgrounds 
(e.g. \cite{Vassilevich:2003xt}). Since the theory we are dealing with is not
renormalisable, the Lagrangian has an infinite series of terms which has to be truncated at
some inverse power of the the cutoff scale of the theory, which we naturally take to be the Planck
mass. At one loop, the $b_2$ coefficient gives us terms up to order $M_p^{-4}$.

A change of variable from $\mu_R$ to $t=\ln(\phi/\mu_R)$ changes the functional form
of the couplings in the effective action from $g_i(\mu_R)$ to $g_i'(t)$,
\begin{equation} 
\Gamma'(g'_i(t),\phi'(t),g'_{\mu\nu}(t),t)=
\Gamma(g_i(\mu_R),\phi(\mu_R),g_{\mu\nu}(\mu_R),\mu_R).\label{running}
\end{equation}
The renormalisation group corrected Lagrangian is defined by the leading term,
${\cal L}'(g'_i(t),\phi'(t),g'_{\mu\nu}(t))$. The dependence of the 
parameters on the Higgs field on $\phi$ is determined by the renormalisation group, which implies
\begin{equation}
{d g'_i\over dt}={\beta_i(g'_j)\over 1+\gamma_\phi(g'_j)},
\end{equation}
subject to values fixed at a given (low energy) mass scale $M$.

The first thing to note about the Coleman-Weinberg method for calculating the $\beta$ 
functions is that it relies on the functional form of the effective action. Therefore a knowledge of
the effective action which is only valid for solutions to the background field equations is not sufficient.
The covariant effective action gives us an unambiguous off-shell formulation and
a unique set of beta functions. In order to construct this covariant effective action we make use of the 
non-trivial  geometry of the space of metrics and fields. In the general case of a gauge 
theory with fields $\varphi^I$ and action $S[\varphi]$, the covariant operator $\Delta_{IJ}$ for the
field fluctuations is given by \cite{Vilkovisky:1984,DeWittdynamical}
\begin{equation}
\Delta_{IJ}=-{\delta^2 S\over\delta\varphi^I\delta\varphi^J}
+\Gamma^K{}_{IJ}{\delta S\over\delta\varphi^K}
+\lambda_gR^{I\alpha}[\varphi]R_{I\alpha}[\varphi],
\label{vdw}
\end{equation}
The innovation of Vilkovisky and DeWitt was to put the second functional derivatives into covariant form
by introducing a field-space connection $\nabla_I$ with connection 
coefficients $\Gamma^K{}_{IJ}$. The connection ensures that the effective action is 
covariant under field redefinitions.  In the Landau gauge limit $\lambda_g\to\infty$, the connection 
coefficients reduce to the Levi-Civita connection
coefficients for the local metric on the space of fields. 
The final term in (\ref{vdw}) is a gauge-fixing term. Details of the covariant approach are given
in Sect \ref{sectcov}

The connection term vanishes when the background field satisfies the classical field equations i.e. 
$\delta S/\delta\varphi^I=0$, and then non-covariant and covariant effective actions agree.
However, we might expect differing results when the background satisfies the 
{\it quantum corrected} field equations. The beta-functions and the renormalisation group corrected 
effective Lagrangians defined using non-covariant and covariant approaches need not be the same. 

In Sects. \ref{grhiggs} and \ref{grmodes} we will calculate the Higgs parameter beta-functions in 
both covariant and non-covariant form in the Einstein and Jordan frames. As is well known, the beta-function
for the curvature coupling $\beta_\xi\propto 6\xi-1$ in the Jordan frame. This standard result
cannot hold in a covariant approach, because $\xi$ vanishes in the Einstein frame and therefore
the covariant $\beta_\xi$ cannot depend on $\xi$. \footnote{In fact, $\xi$ appears as a correction to the
mass in the Einstein frame, and contributes to $\beta_{\mu^2}$.}  As expected, when we do the calculation,
the non-covariant results are frame dependent whilst the covariant results are frame independent. 
However, the combination
$\mu^2+\xi R$, which acts as an effective Higgs mass, and the Higgs self-coupling $\lambda$
have the same scale behaviour in non-covariant and covariant approaches. The leading behaviour
of the renormalisation group effective potential is therefore frame independent, and
differences arise only in terms that are suppressed by factors of $M_p^{-4}$.
Our results for stability with the renormalisation group corrected potential are therefore similar to 
those found previously \cite{Herranen:2014cua}.

In Sect. \ref{grhiggs} we will explore some of the consequences of the covariant approach.
One of these is that field redefinitions mix some of the parameters of the theory,
and having covariance leads to a set of relations between the beta functions for these
parameters. These relations can be used, 
for example, to completely determine the dependence of the beta-functions on the curvature 
coupling $\xi$. Another consequence of using a covariant approach is that the path integral is 
independent of the gauge-fixing terms in the Lagrangian. Therefore quantum tunnelling rates will
be unambiguous. In non-covariant approaches, this issue is non-trivial, and
independence has only been demonstrated explicitly when the true vacuum is not
radiatively generated \cite{Plascencia:2015pga,Endo:2017gal}.

In Sect. \ref{tunnelling}, we look at a practical application of our results to
the decay of the Higgs vacuum during inflation.  The running couplings make
a considerable difference to the Higgs decay rates when the curvature coupling is
small. This is consistent with earlier work by Herranen et al. \cite{Herranen:2014cua}, but we
give more precise results and we confirm that the results are the same in both
the Jordan and Einstein frame. We also find a regime in which the Higgs
potential has two maxima. We only consider vacuum decay using tunnelling 
with the simplest type of instanton, thought recent work has shown the existence of more
complicated instanton solutions \cite{Rajantie:2017ajw}.

This paper focuses on the UV behaviour of the quantum theory and how this affects the
Higgs potential in de Sitter space. In many ways, though, the IR behaviour of Higgs fields 
in de Sitter space is a more interesting subject. It has become apparent, initially from stochastic
theory \cite{Starobinsky:1994bd,Burgess:2009bs,Garbrecht:2011gu,Serreau:2011fu} 
and also from infra-red expansions 
\cite{Rajaraman:2010xd,Beneke:2012kn}, 
that a self-coupled massless scalar field in a de Sitter invariant state 
acquires a mass squared of order $\lambda^{1/2} H^2$, where $H$ is the expansion rate.
This limits the applicability of our results for small curvature coupling. 
It also means that, when integrating the renormalisation group equations for the 
effective mass in de Sitter space, we start with this IR mass, rather than the 
low energy Higgs mass.

\section{Covariant effective actions}\label{sectcov}

The aim of this section is to introduce the field-covariant effective action and to 
give two methods for evaluating the action to one loop order, specifically by taking
the Landau gauge-fixing limit and by decomposition into gauge-fixed and pure 
gauge modes, leading to the results quoted in the introduction.
Field components are denoted by indices $I,J,\dots$ and gauge parameters by 
indices $\alpha,\beta,\dots$.  Condensed notation is used throughout, with contractions 
over $I,J,\dots$ denoting integration over spacetime and functional derivatives
with respect to the fields denoted by $\partial_I$.

A field-covariant effective action can be constructed whenever there exists a covariant notion
of the distance between two field configurations. Formally, this means we have
a Riemannian  geometry on the space of fields $\varphi^I$ and geodesics can be
defined \cite{Vilkovisky:1984,DeWittdynamical,ParkerTomsbook}. 
This geometry allows us to replace an ordinary field displacement 
$\varphi^I-\phi^I$ with the covariant tangent vector to the geodesic from $\phi^I$ to $\varphi^I$, which we denote by 
$\sigma^I(\phi,\varphi)$. The metric can also be used to define a field-space invariant volume measure
$D\varphi$ for functional integration. 

Our starting point is the covariant action of Burgess and Kunstatter 
\cite{Burgess:1987zi,ParkerTomsbook}, defined 
implicitly by,
\begin{equation}
e^{i\Gamma[\phi,\phi_*]}=
\int D\varphi\,e^{iS[\varphi]-i(\delta\Gamma/\delta\sigma^J)(\sigma^J[\phi_*,\phi]-\sigma^J[\phi_*,\varphi])}.
\end{equation}
This depends on the effective field $\phi^I$ and an arbitrary expansion point $\phi_*^I$. The effective
action generates effective field equations for $\phi^I$, in the the sense that
\begin{equation}
\partial_I\Gamma[\phi,\phi_*]=0\implies \langle \sigma[\phi,\hat\varphi]\rangle=0,
\end{equation}
where $\sigma[\phi,\hat\varphi]$ is the geodesic distance and $\hat\varphi^I$ is the field operator.
In the covariant approach, $\phi^I\ne\langle\hat\varphi^I\rangle$, but instead $\phi^I$ is the
classical field which is closest to the quantum field using the invariant distance.
Note that the effective field equations do not depend on the expansion point $\phi_*$. We make
use of this fact and choose $\phi^I_*=\phi^I$, which defines the DeWitt effective action 
\cite{DeWitt2,ParkerTomsbook},
\begin{equation}
\Gamma[\phi]=\Gamma[\phi,\phi]
\end{equation}
The DeWitt effective action generates the effective field equations using $\partial_I\Gamma[\phi]=0$.

In a gauge theory, there are infinitesimal gauge transformations of the field $\varphi^I$ of the form
\begin{equation}
\delta\varphi^I=R^I{}_\alpha\epsilon^\alpha,
\end{equation}
which leave the action invariant, i.e. $R^I{}_\alpha\partial_IS=0$. The gauge is fixed using a gauge-fixing
functional $\chi^\alpha[\phi,\varphi]$, and then the path integral is modified as follows,
\begin{equation}
S[\varphi]\to S[\varphi]+\frac12\gamma_{\alpha\beta}\chi^\alpha\chi^\beta+{\hbar\over i}{\rm tr}\ln\,Q^\alpha{}_\beta,
\end{equation}
This introduces a metric $\gamma_{\alpha\beta}$ on the gauge parameters and a ghost operator
\begin{equation}
Q^\alpha{}_\beta=(\partial_I\chi^\alpha) R^I{}_\beta.
\end{equation}
Next, a procedure developed by Vilkovisky and DeWitt \cite{Vilkovisky:1984} generates the geometry on field space
which guarantees that the effective action is:
\begin{enumerate}
\item Covariant under field redefinitions of $\varphi^I$;
\item Independent of the choice of gauge fixing functional $\chi^\alpha$;
\item Independent of the metric $\gamma_{\alpha\beta}$.
\end{enumerate}

The field-space geometry includes a local field-space metric ${\cal G}_{IJ}$ and a 
non-local field space connection $\nabla_I$. The metric allows an orthogonal decomposition 
of field variations into pure gauge and gauge-fixed directions.
Projection in the pure-gauge direction can be done using
\begin{equation}
\overline\Pi^I{}_J=R^I{}_\alpha{\cal N}^{\alpha\beta}R_{J\beta},
\end{equation}
where indices are lowered using the metric tensors in the usual way, and the normalisation factor appearing here is
\begin{equation}
{\cal N}^{\alpha\beta}=(R^I{}_\alpha R_{I\beta})^{-1}.
\end{equation}
When this is applied to a gauge variation, $\overline\Pi^I{}_J\delta\varphi^J=\delta\varphi^I$.
The orthogonal projection in the gauge-fixed direction is the DeWitt projection $\Pi=\mathbb{I}-\overline\Pi$. 

The local metric also generates a Levy-Civita connection on field space, denoted by ${\cal D}_I$, for example
\begin{equation}
{\cal D}_I{\cal D}_JS=\partial_I\partial_J S-\Gamma^K{}_{IJ} \partial_KS\label{lc}.
\end{equation}
In a gauge theory, the Vilkovisky-DeWitt connection $\nabla_I$ is {\it not} equal to the Levy-Civita connection, 
but it is related to it by the projection operators,
\begin{equation}
\nabla^I\nabla_JS=\Pi({\cal D}^I{\cal D}_JS)\Pi.
\end{equation}
One of the disadvantages of using a covariant approach is that this expression is non-local.
However, we will now describe ways to deal with this non-locality at one loop.

At one-loop, the contribution to the covariant effective action obtained from a geodesic expansion
of the fields in the path integral is
\begin{equation}
\Gamma^{(1)}={\hbar\over 2i}{\rm tr}\ln\left\{\Pi({\cal D}^I{\cal D}_JS)\Pi+
(\partial^I\chi^\alpha)(\partial_J\chi_\alpha)\right\}-{\hbar\over i}{\rm tr}\ln Q^\alpha{}_\beta,
\end{equation}
For an actual calculation, we can reduce the amount of work by choosing a convenient
gauge-fixing functional, in particular the ${\cal R}_\xi$ gauges in which
$\partial_I\chi^\alpha=\lambda_g^{1/2}R_I{}^\alpha$, where $\lambda_g$ is a constant gauge-fixing 
parameter. The one-loop effective action is then
\begin{equation}
\Gamma^{(1)}={\hbar\over 2i}{\rm tr}\ln\left\{\Pi({\cal D}^I{\cal D}_JS)\Pi+\lambda_g R^{I\alpha}R_{J\alpha}
\right\}-{\hbar\over i}{\rm tr}\ln \left\{\lambda_g^{1/2}R^{I\alpha}R_{I\beta}\right\}.\label{vdwloop}
\end{equation}
If the covariant derivatives in (\ref{lc}) are replaced by ordinary functional derivatives, and the
projections are dropped, then the result is a non-covariant effective action contribution 
$\Gamma_{\rm nc}^{(1)}$,
\begin{equation}
\Gamma_{\rm nc}^{(1)}={\hbar\over 2i}{\rm tr}\ln\left\{\partial^I\partial_JS+\lambda_g R^{I\alpha}R_{J\alpha}
\right\}-{\hbar\over i}{\rm tr}\ln \left\{\lambda_g^{1/2}R^{I\alpha}R_{I\beta}\right\}.\label{nc}
\end{equation}
If the background fields are `on shell', specifically when $\partial_IS=0$, then the connection
$\nabla_I\to\partial_I$, and the covariant and non-covariant results coincide, 
$\Gamma^{(1)}=\Gamma_{\rm nc}^{(1)}$. Most calculations
are done on shell, and Eq. (\ref{nc}) is the traditional route to evaluation of the effective action.

We will show that the off-shell result can be simplified in two equivalent ways. Firstly
\begin{equation}
\Gamma^{(1)}={\hbar\over 2i}{\rm tr}\ln\left\{\Pi({\cal D}^I{\cal D}_JS)\Pi
\right\}-{\hbar\over 2i}{\rm tr}\ln \left\{R^{I\alpha}R_{I\beta}\right\},\label{physical}
\end{equation}
where the logarithms are interpreted in a particular way described below. If we have $n$ fields and 
$m$ gauge variations, then there are $n-m$ non-gauge fields but there are $n-2m$ degrees of freedom. 
The ghost contribution accounts for the difference between these two.

The second method is to use the Landau gauge $\lambda_g\to\infty$,
\begin{equation}
\Gamma^{(1)}=\lim_{\lambda_g\to\infty}{\hbar\over 2i}{\rm tr}\ln\left\{{\cal D}^I{\cal D}_JS
+\lambda_g R^{I\alpha}R_{J\alpha}
\right\}-{\hbar\over i}{\rm tr}\ln \left\{\lambda_g^{1/2}R^{I\alpha}R_{I\beta}\right\},\label{landau}
\end{equation}
This appears to be more complicated, but the advantage of this method is that removing the
projection operators leaves an operator which is explicitly local in spacetime,
making it suitable for adiabatic expansion techniques.

For simplicity, we define the functional traces using Euclidean methods with
\begin{equation}
{\rm tr}\ln\,A_L=-i\zeta'(0,A)-i\zeta(0,A)\ln \mu_R^2,
\end{equation}
where $A$ is a positive definite operator obtained from the Lorentzian operator 
$A_L$ by analytic continuation of the time-like coordinate.  This limits us to metrics with a valid analytic
continuation. The generalised zeta-function is defined by $\zeta(s,A)={\rm tr}A^{-s}$ 
and  $\mu_R$ is the renormalisation scale.
We can read off the scaling of the Euclidean effective action $\Gamma_E$ from (\ref{landau})
\begin{equation}
\mu_R{d\Gamma^{(1)}_E\over d\mu_R}=\hbar\lim_{\lambda_g\to\infty}\left\{
-\zeta\left(0,{\cal D}^I{\cal D}_J S+\lambda_gR^{I\alpha}R_{J\alpha}\right)
+2\zeta\left(0,\lambda_g^{1/2}{\cal R}^{I\alpha}{\cal R}_{I\beta}\right)\right\}.\label{escaling}
\end{equation}
In Landau gauge, the operators are local, and it is possible to prove that $\zeta(0,A)$ 
can be expressed in terms of a local adiabatic expansion coefficient $b_2(A)$,
\begin{equation}
\zeta(0,A)={1\over 16\pi^2}\int b_2(A) |g|^{1/2}d^4x.
\end{equation}
Eq. (\ref{escaling}) is the origin of the renormalisation group equation (\ref{scaling}) we gave in the introduction.
For Laplace type operators $A=-\nabla^2+E$, the expansion coefficient $b_2(A)$ is an invariant
polynomial combination of the spacetime curvature and derivatives of $E$. In Ref \cite{Moss:2013cba}, 
it was shown that the expansion coefficients remain polynomial for some classes of non-Laplacian operators
relevant to the covariant effective action. In these cases, we can use $b_2(A)$ to read off
the rescaling behaviour of the terms in the effective potential or the effective Lagrangian
using the renormalisation group equation (\ref{scaling}).

Since the scaling relations obtained from $b_2$ are local, and the difference between covariant and non-covariant 
effective actions vanishes when $\partial_IS=0$, we  can conclude that
\begin{equation}
\mu_R{d\Gamma^{(1)}_{\rm nc}\over d\mu_R}-\mu_R{d\Gamma_E^{(1)}\over d\mu_R}=f^I\partial_IS,
\label{difference}
\end{equation}
where $f^I$ is a tensor polynomial expression of order $\hbar$ in the loop expansion. 
The non-covariant results are mostly known already, so this relation provides both a 
check on new covariant results and a possible route to
finding the covariant $\beta$-functions. Furthermore, we can combine the terms at one loop order
into
\begin{equation}
\mu_R{d\Gamma^{(1)}_{\rm nc}[\varphi]\over d\mu_R}=
\mu_R{d\Gamma_E^{(1)}[\varphi+f\ln\mu_R]\over d\mu_R}+O(\hbar^2).\label{oneloop}
\end{equation}
Note that this does not imply that the approaches are equivalent, because the
covariant action is {\it covariant} not {\it invariant} under field redefinitions.
In the introduction, we described how to construct the renormalisation group corrected
Lagrangian ${\cal L}'$ from the rescaling behaviour of the action at one loop. 
Following the same procedure we see that the effective actions $\Gamma'$ constructed 
from the covariant and the non-covariant effective actions are related by a field transformation,
\begin{equation}
\Gamma'_{\rm nc}[\varphi]=\Gamma_E'[\varphi-ft],
\end{equation}
where $t=\ln(\phi/M)$.
After expanding out again
\begin{equation}
\Gamma'_{\rm nc}[\varphi]=\Gamma_E'[\varphi]-tf^I\,\partial_I\Gamma_E'[\varphi]
+O(t^2\hbar^2).
\end{equation}
At this point, we are unsure about the size of the $O(t^2\hbar^2)$ terms. In the one loop
result (\ref{oneloop}), logarithms endanger the loop expansion when the 
logarithms are large. We corrected for this by using the effective couplings. 
The new remainder term depends on the combination $ft$.
If we take the case where the remainder is small, and apply the corrected field 
equations $\partial_I\Gamma'=0$, as we might use to calculate tunnelling amplitudes 
for example, then the non-covariant action has the same value as the covariant action.
However, we see that for large field values the remainder term may be large,
and there is a possible discrepancy between the non-covariant and covariant 
renormalisation group corrected potentials even when on shell.

To obtain the two representations of the covariant effective action
given earlier, first split $\varphi^I\to(\xi^I,\theta^I)$ into non-gauge and pure-gauge directions. 
Decompose the operator ${\cal D}^I{\cal D}_JS$ as
\begin{equation}
{\cal D}^I{\cal D}_JS=A=
\begin{pmatrix}
a&c\\
c^\dagger&d\\
\end{pmatrix}
\end{equation}
Similarly decompose
\begin{equation}
R^{I\alpha}R_{J\alpha}=B=
\begin{pmatrix}
0&0\\
0&b\\
\end{pmatrix}
\end{equation}
Eq. (\ref{physical}) follows from this decomposition when we set $\lambda_g=1$
in the one-loop result (\ref{vdwloop}). Noting that the non-zero eigenvalues
of $R^{I\alpha}R_{J\alpha}$ and $R^{I\alpha}R_{I\beta}$ are identical,
\begin{equation}
\Gamma^{(1)}={\hbar\over 2i}{\rm tr}\ln a+{\hbar\over 2i}{\rm tr}\ln b
-{\hbar\over i}{\rm tr}\ln b={\hbar\over 2i}{\rm tr}\ln a-{\hbar\over 2i}{\rm tr}\ln b.\label{physred}
\end{equation}
This recovers Eq. (\ref{physical}).

For Landau gauge Eq. (\ref{landau}), we start with the generalised zeta-function $\zeta(s,A)$. Consider
\begin{equation}
\zeta(s,A+\lambda_g B)=
{1\over\Gamma(s)}\int_0^\infty dt\,t^{s-1}{\rm tr}\left(e^{-(A+\lambda_g B)t}\right)
\end{equation}
If we rescale $t$,
\begin{equation}
\zeta(s,A+\lambda_g B)=
{\lambda_g^{-s}\over\Gamma(s)}\int_0^\infty dt\, t^{s-1}{\rm tr}\left(e^{-(B+\lambda_g^{-1} A)t}\right)
\end{equation}
Separate out the diagonal and non-diagonal parts,
\begin{equation}
{\rm tr}\left(e^{-(B+\lambda_g^{-1} A)t}\right)=
{\rm tr}\left(
\exp\left[-
\begin{pmatrix}
\lambda_g^{-1}a&0\\
0&b+\lambda_g^{-1}d\\
\end{pmatrix}t
\right]\exp\left[-
\begin{pmatrix}
0&\lambda_g^{-1}c\\
\lambda_g^{-1}c^\dagger&0\\
\end{pmatrix}t
\right]\right)
\end{equation}
Only the even powers of $\lambda_g^{-1}$ survive in the second exponential due to the trace.
Of these, only the leading term survives in the large $\lambda_g$ limit, and after rescaling $t$
back,
\begin{equation}
\zeta(s,A+\lambda_g B)=\zeta(s,a)+\lambda_g^{-s}\zeta(s,b+\lambda_g^{-1}d)+O(\lambda_g^{-s-2})
\end{equation}
We use analytic continuation to $s=0$ and then the limit $\lambda_g\to \infty$,
\begin{eqnarray}
\zeta(0,A+\lambda_g B)&\sim&\zeta(0,a)+\zeta(0,b)\\
\zeta'(0,A+\lambda_g B)&\sim&\zeta'(0,a)+\zeta'(0,b)-\zeta(0,b)\ln\lambda_g
\end{eqnarray}
So now the terms on the right hand side of (\ref{landau}) are
\begin{equation}
\lim_{\lambda_g\to\infty}\frac{\hbar}{2i}{\rm tr}\ln\left\{A+\lambda_g B
\right\}-{\hbar\over i}{\rm tr}\ln \left\{\lambda_g^{1/2}b\right\}=
\frac{\hbar}{2i}{\rm tr}\ln a-\frac{\hbar}{2i}{\rm tr}\ln b
\end{equation}
Therefore the Landau gauge result Eq. (\ref{landau}) is equal to Eq. (\ref{physred}) which
is equal to the gauge decomposition Eq. (\ref{physical}).

\section{The Gravity-Higgs effective field theory}\label{ghtheory}

We take the point of view that the gravity-Higgs sector is a low energy
effective field theory for the spacetime metric $g_{\mu\nu}$ and the Higgs doublet field ${\cal H}$,
in which non-renormalisable terms are assumed to be suppressed by inverse powers of the
reduced Planck mass, $\kappa=M_p^{-1}=(8\pi G)^{1/2}$ \cite{Burgess:2003jk}.
Since Higgs instability sets in at a scale below the Planck mass, the renormalisable
couplings will be expected to play the most important role. 
During inflation, we suppose that the vacuum energy is dominated by an inflation
field and takes some fixed value $V_0$, and then the expansion rate in the Higgs vacuum 
is determined by the Friedman equation $H^2=\kappa^2 V_0/3$.

For convenience, we replace the Higgs doublet by a set of four real scalars $\phi^i$,
denoting the gauge invariant magnitude of the field by $\phi$ and the projection
orthogonal to $\phi^i$ by $\delta^\perp_{ij}$. The Lagrangian 
density for the gravity-Higgs sector ${\cal L}_g$ with non-minimal coupling is
\begin{equation}
{\cal L}_g(g,\phi)={1\over 2\kappa^2}U(\phi)R(g)\,|g|^{1/2}
-\frac12 G_{ij}(\phi)\,g^{\mu\nu}\partial_\mu\phi^i\partial_\nu\phi^j\,|g|^{1/2}
-V(\phi)\,|g|^{1/2},\label{nlsm}
\end{equation}
where $\partial_\mu$ denotes an ordinary spacetime derivative. The non-minimal coupling terms are contained 
in the function $U(\phi)$ multiplying the Ricci scalar $R$. 

Each one of the scalar functions in the Lagrangian has an expansion in powers of $\kappa$,
\begin{eqnarray}
V(\phi)&=&V_0+\frac12\mu^2\phi^2+\frac14\lambda\phi^4+\frac16\lambda_6\kappa^2\phi^6
+\dots\label{vexpand}\\
G_{ij}(\phi)&=&\delta_{ij}+\alpha\kappa^2\delta_{ik}\delta_{jl}\phi^k\phi^l+
\beta\kappa^2\delta^\perp_{ij}\phi^2+\dots\\
U(\phi)&=&1-\xi\kappa^2\phi^2+\dots,\label{uexpand}
\end{eqnarray}
Most of the results we obtain have been truncated to $O(\kappa^2\phi^2)$. We will use a wave
function renormalisation to the keep the leading order behaviour in $R$ and $G_{ij}$ fixed.
The anomalous dimensions will be denoted by $\gamma_g$ and $\gamma_\phi$
respectivly. This keeps the effective Planck scale fixed.
Note that it is not possible to eliminate both coefficients $\alpha$ and $\beta$
by redefinitions of $\phi$ if $G_{ij}$ has a non-vanishing curvature tensor.

One of the questions we address is the effect of conformal rescaling of the metric from the
original Jordan Frame to the Einstein frame to remove the $\xi$ term in the original Lagrangian.
We set the metric to be $g_E=U(\phi)g$, then
\begin{equation}
{\cal L}_g(g_E,\phi)={1\over 2\kappa^2}R(g_E)\,|g_E|^{1/2}
-\frac12 G_{Eij}(\phi)\,g_E^{\mu\nu}\partial_\mu\phi^i\partial_\nu\phi^j\,|g_E|^{1/2}
-V_E(\phi)\,|g_E|^{1/2},\label{nlsm}
\end{equation}
where
\begin{eqnarray}
V_E(\phi)&=&U^{-2}V(\phi)\\
G_{Eij}(\phi)&=&U^{-1}G_{ij}+\frac32\kappa^{-2}U^{-2}
{\partial U\over\partial \phi^i}{\partial U\over\partial \phi^j}
\end{eqnarray}
In a covariant theory it should be possible to calculate the beta functions by transforming to the
Einstein frame, rescaling the effective action, and transforming back to the Jordan frame.

If we expand the Einstein frame theory in powers of $\kappa$ we have relationships between
the sets of Einstein frame and Jordan frame parameters,
\begin{eqnarray}
\mu_E^2&=&\mu^2+4\xi'\kappa^2 V_0\\
\lambda_E&=&\lambda+4\xi'\kappa^2\mu^2\\
\xi_E&=&\xi-\xi'.
\end{eqnarray}
Note that we have used a different $\xi'$ for the conformal transformation. Since we have
adopted a covariant quantisation approach, these relations also hold for the running couplings
up to field renormalisation factors. We differentiate the relations with respect to the renormalisation 
scale keeping $\xi'$ fixed, and then set $\xi'=\xi$ at the end,
\begin{eqnarray}
\tilde\beta_{\mu^2}(0,\lambda_E,\mu^2_E,\dots)&=&\tilde\beta_{\mu^2}(\xi,\lambda,\mu^2,\dots)
+4\xi\kappa^2 \tilde\beta_{V_0}(\xi,\lambda,\mu^2,\dots)\label{betaI}\\
\tilde\beta_\lambda(0,\lambda_E,\mu^2_E,\dots)&=&\tilde\beta_\lambda(\xi,\lambda,\mu^2,\dots)
+4\xi\kappa^2\tilde\beta_{\mu^2}(\xi,\lambda,\mu^2,\dots)\\
\tilde\beta_\xi(0,\lambda_E,\mu^2_E,\dots)&=&\tilde\beta_\xi(\xi,\lambda,\mu^2,\dots)\label{betaIII}
\end{eqnarray}
The beta functions $\tilde\beta$ include anomalous dimension factors, for example
\begin{equation}
\tilde\beta_\xi=\beta_\xi-2\gamma_\phi\xi-\gamma_g\xi.
\end{equation}
These relations can be used to evaluate covariant beta functions for non-zero curvature
coupling if we have results for minimal coupling.

Already, an unexpected result follows from (\ref{betaIII}), namely that the one-loop $\beta_\xi$ 
for gravity-Higgs theory is independent of $\xi$ at order $\kappa^0$. Paradoxically, the quantum theory 
of scalar fields on a curved background gives $\beta_\xi\propto 6\xi-1$ \cite{Buchbinder:1986yh}. 
The $\xi$ dependence 
must cancel when we include quantum gravity and require field-covariance of the effective action. 
Subsequent results will confirm this using explicit calculations.

\section{Expansions of the gravity-Higgs action}\label{grhiggs}

We will give results for the second order variations of the gravity-Higgs effective theory
which are needed to evaluate the beta functions. Most of the details have been left out 
because these are already covered in the literature, particularly in the
work of Barvinsky et al.  \cite{Barvinsky:2009ii,Barvinsky:2009fy,Barvinsky:2010yb}). 
We use the Jordan frame formulation
and then the covariant formulation can be checked by verifying the relations between the
beta functions give in Eqs (\ref{betaI})-(\ref{betaIII}).

Variations in field space can be combined into metric and scalar directions,
and we rescale these to have the same dimensions, i.e.
\begin{equation}
\partial_IS=\left(2\kappa{\delta S\over \delta g_{\mu\nu}},{\delta S\over\delta\phi^i}\right)
\end{equation}
The second-order variation of the action gives a second order differential operator,
\begin{equation}
-{\cal D}_I{\cal D}_JS=-\partial_I\partial_JS+
\Gamma^K{}_{IJ}\partial_KS=
-{\cal G}_{IJ}\nabla^2-{\cal P}^{\alpha\beta}{}_{IJ}\nabla_\alpha\nabla_\beta
+E_{IJ},\label{soo}
\end{equation}
There are three important tensors in this expression
which will be presented below: ${\cal G}_{IJ}$  is the metric on field space, ${\cal P}^{\alpha\beta}_{IJ}$ 
projects out the gauge-fixed directions, and $E_{IJ}$ is an effective mass term. The metric is used to construct the
Levy-Civita connection by the usual expression,
\begin{equation}
\Gamma^I{}_{JK}=\frac12 {\cal G}^{IL}\left(\partial_K{\cal G}_{LJ}
+\partial_J{\cal G}_{LK}-\partial_L{\cal G}_{JK}\right).\label{lcc}
\end{equation}
When writing down local operators like ${\cal G}_{IJ}$ we usually omit delta function terms.

\subsection{First order variations}

The first order variation of the action defines the background field equations for the gravitational
and Higgs fields, which will be denoted by $F^{\mu\nu}$ and $F_i$,
\begin{align}
&F^{\mu\nu}=-{\kappa\over\sqrt{g}}{\delta S\over\delta g_{\mu\nu}}=
UG^{\mu\nu}-U^{;\mu\nu}+g^{\mu\nu}U^{;\rho}{}_\rho-\kappa^2T^{\mu\nu},\label{einsteineq}\\
&F_i=-{1\over \sqrt{g}}{\delta S\over\delta \phi^i}=
-{1\over 2\kappa^2}RU_{,i}-D_\mu (G_{ij}\nabla^\mu \phi^i)+V_{,i},\label{scalareq}
\end{align}
where $D_\mu$ is a covariant derivative for the metric $G_{ij}$,
\begin{equation}
D_\mu\delta\phi^i=\partial_\mu\delta\phi^i-\Gamma^i{}_{jk}(\partial_\mu\phi^j)\delta\phi^k
\end{equation}
Note that $F^{\mu\nu}$ has been scaled so that $F^{\mu\nu}=0$ resembles the usual Einstein equation.

\subsection{Second order variations}

For simplicity, {\it the background scalar field will be assumed constant}.
The second order variation $-\partial_I\partial_JS_g$ has derivative terms
\begin{equation}
\begin{pmatrix}
-Ug^{(\mu\nu)(\rho\sigma)}\nabla^2+UP^{\alpha\beta(\mu\nu)(\rho\sigma)}\nabla_\alpha\nabla_\beta
&-\kappa^{-1}U_{,j}(\nabla^\mu\nabla^\nu-g^{\mu\nu}\nabla^2)\\
-\kappa^{-1}U_{,i}(\nabla^\rho\nabla^\sigma-g^{\rho\sigma}\nabla^2)&
-G_{ij}\nabla^2\\
\end{pmatrix}|g|^{1/2},
\label{gmass}
\end{equation}
and a potential term,
\begin{equation}
E_{g\,IJ}=
\begin{pmatrix}
E_g^{(\mu\nu)(\rho\sigma)}
&\kappa g^{\mu\nu}V_{,j}\\
\kappa g^{\rho\sigma} V_{,i}&
V_{,ij}-\frac12R\kappa^{-2}U_{,ij}\\
\end{pmatrix}|g|^{1/2},
\label{gpot}
\end{equation}
Two important tensors in the kinetic terms are the DeWitt metric,
\begin{equation}
g^{(\mu\nu)(\rho\sigma)}=
\frac12(g^{\mu\rho}g^{\nu\sigma}+g^{\mu\sigma}g^{\nu\rho}-g^{\mu\nu}g^{\rho\sigma}),
\end{equation}
and another tensor which will also appear in the gauge-fixing terms below,
\begin{equation}
P^{\alpha\beta(\mu\nu)(\rho\sigma)}=2g_{\gamma\delta}g^{(\alpha\gamma)(\mu\nu)}g^{(\beta\delta)(\rho\sigma)}.
\end{equation}
The mass-like gravity terms are
\begin{equation}
E_g^{(\mu\nu)(\rho\sigma)}=-2UR^{\dot\mu\rho\dot\nu\sigma}+2UR^{\dot\mu\rho}g^{\dot\nu\sigma}
+UR_T^{\mu\nu}g^{\rho\sigma}+Ug^{\mu\nu}R_T^{\rho\sigma}
-4UR_T^{\dot\mu\rho}g^{\dot\nu\sigma}-2\kappa^2 Vg^{(\mu\nu)(\rho\sigma)}.\label{gmass}
\end{equation}
Dots over indices indicate symmetrisation in those indices and a subscript $T$ denotes the
trace-free part of the tensor. The terms have been organised this way to isolate terms which
vanish when the differential operator is applied to transverse traceless perturbations and terms
which remain. 

Gauge-fixing terms have to be included in the action, and following Barvinski \cite{Barvinsky:2009ii}
we take
\begin{equation}
{\cal L}_{\rm gf}=-\lambda_gUg_{\mu\nu}\chi^\mu\chi^\nu,\label{gf}
\end{equation}
where
\begin{equation}
\chi^\mu={1\over 2\kappa}(g^{(\mu\nu)(\rho\sigma)}\nabla_\nu\delta g_{\rho\sigma}-
U^{-1}U_{,i}\nabla^\mu\delta\phi^i)\label{gft}
\end{equation}
The gauge-fixing term gives a contribution to the second-order variation of,
\begin{equation}
-\partial_I\partial_J S_{gf}=-\lambda_g
\begin{pmatrix}
UP^{\alpha\beta(\mu\nu)(\rho\sigma)}\nabla_\alpha\nabla_\beta&
-\kappa^{-1}U_{,j}g^{(\mu\nu)(\alpha\beta)}\nabla_\alpha\nabla_\beta\\
-\kappa^{-1}U_{,i}g^{(\rho\sigma)(\alpha\beta)}\nabla_\alpha\nabla_\beta&
\frac12\kappa^{-2}U^{-1}U_{,i}U_{,j}\nabla^2\\
\end{pmatrix}|g|^{1/2},
\end{equation}
We have enough information now to obtain the field-space metric ${\cal G}_{IJ}$. We will do this by
requiring the operator to have Laplacian form in the gauge-fixed directions, i.e. 
\begin{equation}
-(\partial_I\partial_JS_g)\delta\varphi^I=-{\cal G}_{IJ}\nabla^2\delta\varphi^J+E_{IJ}\delta\varphi^J
\end{equation}
when $\chi^\mu(\delta\varphi)=0$. Note that variations of the gauge-fixing term vanish 
when applied to the gauge-fixed directions, and so an arbitrary amount
of $\partial_I\partial_JS_{gf}$ can be added to the differential operator. However,
$(\partial_I\partial_JS_g)+\lambda_g^{-1} (\partial_I\partial_JS_{gf})$ 
is the unique combination of the second order variations that has Laplacian form. 
The coefficient of $-\nabla^2$ in this combination is therefore the field-space metric, and this gives
\begin{equation}
{\cal G}_{IJ}=
\begin{pmatrix}
Ug^{(\mu\nu)(\rho\sigma)}&
-\frac12\kappa^{-1}U_{,j}g^{\mu\nu}\\
-\frac12\kappa^{-1}U_{,i}g^{\rho\sigma}&
G_{ij}+\frac12\kappa^{-2}U^{-1}U_{,i}U_{,j}\\
\end{pmatrix}|g|^{1/2}.
\end{equation}
(We also obtain exactly the same result using $\partial_I\chi^\alpha=\lambda_g^{1/2}R_I{}^\alpha$ for 
the $R_\xi$ gauges as in Sect. \ref{sectcov}. This is
a special feature of the gauge fixing term (\ref{gft}), and for other choices it 
becomes necessary to combine information from both gauge-fixed and pure gauge 
directions to obtain the metric).
For future reference, the inverse metric is given by
\begin{equation}
{\cal G}^{IJ}=
\begin{pmatrix}
U^{-1}g_{(\mu\nu)(\rho\sigma)}+\frac14\kappa^{-2}U^{-1}U_{,k}U^{,k}W^{-1}g_{\mu\nu}g_{\rho\sigma}&
-\frac12\kappa^{-1}W^{-1}U^{,j}g_{\mu\nu}\\
-\frac12\kappa^{-1}W^{-1}U^{,i}g_{\rho\sigma}&
G^{ij}-\frac32W^{-1}\kappa^{-2}U^{,i}U^{,j}\\
\end{pmatrix}|g|^{-1/2}.
\end{equation}
where $W=U+\frac32\kappa^{-2}U^{-1}U_{,i}U^{,i}$.

Finally, comparing to the expression (\ref{soo}), we can also read off the tensor 
${\cal P}^{\alpha\beta}{}_{IJ}$,
\begin{equation}
{\cal P}^{\alpha\beta}{}_{IJ}=
\begin{pmatrix}
UP^{\alpha\beta(\mu\nu)(\rho\sigma)}&
-\kappa^{-1}U_{,j}g^{(\mu\nu)(\alpha\beta)}\\
-\kappa^{-1}U_{,i}g^{(\rho\sigma)(\alpha\beta)}&
\frac12\kappa^{-2}U^{-1}U_{,i}U_{,j}\\
\end{pmatrix}|g|^{1/2}.\label{pdef}
\end{equation}
The variations are considerably simpler in the Einstein frame $U\equiv1$. Indeed, one of the
motivations for considering covariant approaches is to ensure that the Einstein frame result
can always be used reliably. However, just this once, we are going to verify that the covariant
result is independent of the choice of conformal frame.

\subsection{Vilkovisky-DeWitt corrections}

The Levy-Civita connection is given by the usual expression (\ref{lcc}).
The connection converts the scalar derivatives $V_{,ij}$ into covariant derivatives $V_{;ij}$,
and adds extra terms to the differential operator (\ref{soo}). For simplicity, we just quote the contributions 
up to $O(\kappa^2)$, and take the spacetime curvature to be of order $\kappa^2$ and 
the scalar derivatives $V_{,i}$ of order $\kappa$, then
\begin{equation}
\Gamma^K{}_{IJ}\partial_KS_g=
\begin{pmatrix}
E_\Gamma{}^{(\mu\nu)(\mu\nu)}&-\frac14\kappa (2V_{,j}-\kappa^{-2}RU_{,j})g^{\mu\nu}\\
-\frac14\kappa (2V_{,i}-\kappa^{-2}RU_{,i})g^{\rho\sigma}&\frac12(G_{ij}+\kappa^{-2}U_{;ij})(R-4\kappa^2V)\\
\end{pmatrix},\label{vdwco}
\end{equation}
where
\begin{equation}
E_\Gamma{}^{(\mu\nu)(\rho\sigma)}=
2Ug^{\dot\mu\rho}R_T^{\dot\nu\sigma}-\frac12UR_T^{\mu\nu}g^{\rho\sigma}-\frac12Ug^{\rho\sigma}R_T^{\mu\nu}.
\end{equation}
Eq. (\ref{vdwco}) is exact when in the minimally coupled case $U=1$.
In the non-minimally coupled case, when the  Levy-Civita connection term is combined with the 
mass terms from the second variation of the action (\ref{gpot}), we see that the curvature coupling 
terms $U_{;ij}R$ cancel. In particular, the $2\xi\kappa^2 R$ term which would contribute $\xi$ dependence
to the beta-function $\beta_\xi$ has been cancelled off by the Vilkovisky-DeWitt corrections.

\section{Gravity-Higgs mode expansions}\label{grmodes}

We argued in Sect \ref{sectcov} that the scaling behaviour of the gravity-scalar effective action can be 
expressed in terms of spacetime invariant tensor combinations. General expressions for these
combinations are known from heat kernel 
methods for a wide range of second order operators \cite{Vassilevich:2003xt,Moss:2013cba}, 
but there are some non-Laplace type 
operators where the general results are not yet available. Furthermore, it can be very cumbersome
applying these general results. A more practical approach is to use a direct 
evaluation of the the generalised zeta function on a simple manifold for a simple operator, 
for example gravity with a single scalar field on the sphere \cite{Dowker:1993ww,Cognola:2005de}, 
and read off the relevant coefficients.

On the sphere $S_4$, the curvature is given in terms of the Ricci scalar
\begin{equation}
R_{\mu\nu\rho\sigma}=\frac1{12}R\left(g_{\mu\rho}g_{\nu\sigma}-g_{\mu\sigma}g_{\nu\rho}\right),
\end{equation}
and the radius of the sphere is $\sqrt{12/R}$.
Consider a single constant scalar field $\phi$ with Euclidean Lagrangian
\begin{equation}
{\cal L}_E=\frac12 K(\phi)(\nabla\phi)^2+V(\phi)-{1\over\kappa^2}U(\phi)R
\end{equation}
The second order operator for the Euclidean theory is
\begin{equation}
{\cal D}^I{\cal D}_JS_E=-\delta^I{}_J\nabla^2+(\lambda_g-1){\cal R}^{I\alpha}{\cal R}_{J\alpha}+E^I{}_J,
\end{equation}
where the general expressions for ${\cal R}^{I\alpha}{\cal R}_{J\alpha}$ and $E^I{}_J$ 
where given in Sect. \ref{grhiggs}. For a single field, we replace $V_{,i}$ by $V'$ and $V_{;ij}$
by the covariant derivative with metric $K(\phi)$,
\begin{equation}
V''=K^{1/2}(K^{-1/2}V')'.
\end{equation}
The differential operators can be diagonalised by expanding the fields in a basis of $S_4$ orthonormal harmonics:  
scalar modes $h^S$, transverse vector modes $h^V{}_\mu$ and transverse-traceless tensor modes $h^{T}{}_{\mu\nu}$. 
Transverse modes are divergence free, i.e. $\nabla^\mu h^V{}_\mu=0$. The eigenvalues of $-\nabla^2$ 
for the respective modes are $\lambda_S$, $\lambda_V$ and $\lambda_T$. Modes can be traded up 
into higher rank tensors by applying derivatives to the basic set of harmonics.
The general decomposition of the metric plus scalar field into the basis of harmonic functions and their
derivatives is given by mode sums with coefficients $x^I$,
\begin{equation}
\delta g_{\mu\nu}=2\kappa \sum_{\rm modes}\,\left\{x^1h^{T}_{\mu\nu}+2x^2\nabla_{(\mu} h^V{}_{\nu)}
+x^3\nabla_{\mu\nu}h^{S}+x^4g_{\mu\nu}h^S\right\},
\qquad \delta\phi^i=\sum_{\rm modes}\,x^5h^S,
\end{equation}
where $\nabla_{\mu\nu}=\nabla_\mu\nabla_\nu-\frac14g_{\mu\nu}\nabla^2$. In the ghost sector, 
there is a similar decomposition,
\begin{equation}
c_\mu=\sum_{\rm modes}\,\left\{y^1h^V{}_\mu+y^2\nabla_\mu h^S\right\}.
\end{equation}
The eigenvalues of the derived modes change due to non-commutation of the covariant derivatives, for example
\begin{equation}
-\nabla^2\left(\nabla_{(\mu}h^V{}_{\nu)}\right)=
\left(\lambda_V-{\textstyle\frac5{12}}R\right)\nabla_{(\mu}h^V{}_{\nu)}
\end{equation}
The derived modes are not normalised, but their normalisation can be deduced from the
original harmonic, for example
\begin{equation}
4\int \nabla_{(\mu}h^V{}_{\nu)}\nabla^{(\mu}h^{V\,\nu)}|g|^{1/2}d^4x=
2\left(\lambda_V-{\textstyle\frac14}R\right).
\end{equation}
The action of the operators and products in the harmonic basis for a 
given set of eigenvalues can be represented now by $5\times 5$ matrices, 
which are given in appendix \ref{matrices}.

At this point, we would like to stress an important point that
the matrices ${\cal D}_I{\cal D}_JS_E$ are not positive definite, so there are 
directions which decrease the
Euclidean action and invalidate the path integral approach. This is the famous conformal 
mode problem of Euclidean quantum gravity.
However, the matrices representing $\Pi{\cal D}_I{\cal D}_JS_E\Pi$, and
${\cal D}_I{\cal D}_JS_E+\lambda_gR_I{}^\alpha R_{J\alpha}$ for sufficiently
large $\lambda_g$, are both positive definite and the path integral can be defined. 
This is the solution to the conformal mode problem of Euclidean quantum gravity referred to in
Ref. \cite{Moss:2013cba}.

The renormalisation scale dependence of the one-loop effective action can be calculated in 
two different ways, and the comparison gives a check on the accuracy of the result. The first way
is by gauge decomposition,
\begin{equation}
\mu_R{d\Gamma^{(1)}_E\over d\mu_R}=-\zeta\left(0,\Pi({\cal D}^I{\cal D}_J S)\Pi\right)
+\zeta\left(0,{\cal R}^{I\alpha}{\cal R}_{I\beta}\right)
\end{equation}
The second version is Landau gauge,
\begin{equation}
\mu_R{d\Gamma^{(1)}_E\over d\mu_R}=\lim_{\lambda_g\to\infty}\left\{
-\zeta\left(0,{\cal D}^I{\cal D}_J S+\lambda_gR^{I\alpha}R_{J\alpha}\right)
+2\zeta\left(0,\lambda_g^{1/2}{\cal R}^{I\alpha}{\cal R}_{I\beta}\right)\right\}
\end{equation}
In each case, the eigenvalues are evaluated by diagonalising the matrices
and the generalised zeta functions are defined for $s>2$ by series,
\begin{equation}
\zeta(s,A)=\sum_\lambda \lambda^{-s}.
\end{equation}
Spherical harmonic eigenvalues are all quadratic polynomials in a single
`angular momentum' index $n$. After diagonalisation, the eigenvalues
are algebraic functions of the spherical harmonic eigenvalues, but
standard zeta-function methods can be modified to analytically continue
and evaluate $\zeta(0,A)$ \cite{Elizalde:1994gf}.

For example, in the transverse-traceless tensor sector $I=J=1$, the eigenvalues are the same
for either method,
\begin{equation}
\lambda=\lambda_T+U^{-1}m_T^2.
\end{equation}
The tensor eigenvalues are given in appendix \ref{az}, and after analytic continuation using (\ref{zetacont}),
\begin{equation}
\zeta(0,({\cal D}^2S)^1{}_1)=-\frac1{18}+20{m_T^2\over UR}+60{m_T^4\over U^2R^2}
\end{equation}
Contributions to the beta-functions from the transverse-traceless tensors can be obtained from
\begin{equation}
\mu_R{d{\cal L}_E^{(1)}\over d\mu_R}=-{b_2\over 16\pi^2},
\end{equation}
where the adiabatic expansion coefficient $b_2$ can be extracted from
\begin{equation}
b_2={16\pi^2\over{\rm Volume}\,S_4}\,\zeta(0,({\cal D}^2S)^1{}_1)={R^2\over 24}\,\zeta(0,({\cal D}^2S)^1{}_1),
\end{equation}
After substituting for $m_T^2$ (see appendix \ref{az}), the tensor mode contribution to $b_2$ becomes
\begin{equation}
b_2={719\over 432}R^2-{25\over 3}{\kappa^2RV\over U}+10{\kappa^4V^2\over U^2}.
\end{equation}
For example, with $U=1-\xi\kappa^2\phi^2$, we have a contribution to $\beta_\xi$ from expanding the
second term in powers of $\kappa$,
\begin{equation}
\beta_\xi-2\gamma_\phi\xi-\gamma_g\xi=2\,{\rm coeff}(b_2,R\phi^2)={50\over 3}\kappa^4 V_0+O(\kappa^6).
\end{equation}

\begin{table}[htb]
\def\arraystretch{1.1}
\begin{tabular}{|lll|}
\hline
&Jordan frame&Einstein frame\\
\hline
$16\pi^2\beta_\xi$&$(6\xi-1) \lambda$&$- \lambda$\\
$16\pi^2\beta_{\mu^2}$&$6\mu^2\lambda$&$6(\mu^2+4\xi\kappa^2 V_0)\lambda$\\
$16\pi^2(\beta_{\mu^2}+4\kappa^2V_0\beta_\xi)$&$6(\mu^2+4\xi\kappa^2V_0)\lambda-4\lambda\kappa^2 V_0$
&$6(\mu^2+4\xi\kappa^2V_0)\lambda-4\lambda\kappa^2 V_0$\\
\hline
\end{tabular}
\begin{tabular}{|ll|}
\hline
&Covariant\\
\hline
$16\pi^2\beta_\xi$&$2\lambda$\\
$16\pi^2\beta_{\mu^2}$
&$6(\mu^2+4\xi\kappa^2V_0)\lambda-12\lambda\kappa^2 V_0$\\
$16\pi^2(\beta_{\mu^2}+4\kappa^2V_0\beta_\xi)$
&$6(\mu^2+4\xi\kappa^2V_0)\lambda-4\lambda\kappa^2 V_0$\\
\hline
\end{tabular}
\caption{\label{table0}$\beta-$functions for the curvature coupling and the mass
of a gravity coupled scalar field at leading order for small $\kappa^4V_0$. 
The Jordan frame result has been calculated directly from the original action. The 
Einstein frame result is obtained by transforming the action to the Einstein frame. 
The covariant result uses a geodesic expansion on field space and is independent 
of the frame used. The renormalisation group flow of $\mu^2+4\kappa^2V_0\xi$ 
is the same for each of these approaches.
}
\end{table}

\begin{table}[htb]
\def\arraystretch{1.1}
\begin{tabular}{|lll|}
\hline
&Jordan frame&Covariant\\
\hline
$16\pi^2\beta_\lambda$&$18\lambda^2$&$18\lambda^2$\\
$16\pi^2\beta_6$&$90\lambda\lambda_6-18\lambda^2(2-8\xi+18\xi^2)$
&$90\lambda\lambda_6-18\lambda^2(1-7\xi+24\xi^2)$\\
$16\pi^2\gamma_g$&$-\frac13\kappa^2\mu^2+2\kappa^2(\mu^2\xi-4\kappa^2V_0)$&
$\frac23(\mu^2+4\xi\kappa^2V_0)\kappa^2-\frac{52}3\kappa^4V_0$\\
\hline
\end{tabular}
\caption{\label{table1}$\beta-$functions for the quartic scalar self-coupling $\lambda$ and the 
sixth order scalar self-coupling $\lambda_6$ of a gravity coupled scalar 
field at leading order for small $\kappa^4V_0$.
The wave function renormalisation of the metric $\gamma_g$ is given at
order $\kappa^4V_0$. The Jordan frame results have been calculated directly from the original action. The 
Einstein frame results are obtained by transforming the action to the Einstein frame. 
The covariant result uses a geodesic expansion on field space and is independent of the frame used.
}
\end{table}

Other contributions to the beta functions can be obtained in a similar way from the matrices
given above, but the details are lengthy and unilluminating. The results given below 
have been obtained using MAPLE and most of them have been checked by hand.
We will give results for the contributions to the beta functions from the Higgs
background direction and the gravitational sector with which it mixes. Other contributions
from the Higgs components perpendicular to the background direction (`Goldstone-like
modes') will be given in a later section. 

Tables \ref{table0} and \ref{table1} show results at leading order for small $\kappa^4V_0$, assuming that the curvature $R$ 
and the mass square $\mu^2$ are of order $\kappa^2V_0$. These choices are consistent when the application 
is to Higgs stability, where the curvature of the universe $R\approx4\kappa^2V_0$ 
and the Higgs mass is negligible. Contributions to the beta functions from the gravitational 
perturbations, like the transverse traceless tensor modes discussed above, enter only at 
order $\kappa^4V_0$, in agreement with the conclusion of Ref. \cite{Markkanen:2017dlc}.
However, quantum gravity does has an effect at leading order through the
Vilkovisky-DeWitt connection terms in the operators. 

The first thing to notice in table \ref{table0}  is the absence of $\xi$ terms for $\beta_\xi$ in the Einstein
frame and the covariant results. The reason for this in the Einstein frame is
obvious, since the $\xi R\phi^2$ term has been eliminated by the conformal
transformation, and $\xi$ appears in the Einstein frame scalar potential
$V_J/U^2$ instead. The absence of $\xi$ in the covariant 
results follows from the beta-function relations (\ref{betaI}-\ref{betaIII}). 
In the explicit calculation, the leading order contribution to $\beta_\xi$ from $RU''$ 
in the mass matrix (\ref{massmat}) cancels with the Vilkovisky-DeWitt correction.

In table \ref{table0} we see that the renormalisation group flow of 
$\mu^2+4\kappa^2V_0\xi$ is the same in all the different approaches. 
For Higgs stability, $R\approx 4\kappa^2 V_0$, and the effective square 
mass of the Higgs field $\mu^2+\xi R\approx \mu^2+4\kappa^2V_0\xi$.
This is the crucial combination which determines whether the Higgs 
can survive in the false vacuum during early universe
inflation, as we shall see later in Sect. \ref{tunnelling}.  The non-covariant
formulation in the Jordan or the Einstein frame therefore gives the same 
outcome for stability as the covariant approach, at least for small
values of $\kappa^4V_0$.

\section{Gauge bosons, Goldstone modes and fermions}

We turn now to gauge bosons, Goldstone modes and fermion contributions to the
effective action for the scalar field on a curved spacetime background. In this case, 
there are no background gauge fields and the gauge modes decouple from the
graviton and scalar modes of the previous section. Quantum gravity still has an effect
via the Vilkovisky-DeWitt connection term. We will describe the calculation in the Einstein frame, 
using the beta-function relations 
(\ref{betaI}-\ref{betaIII}) to extend the results to the Jordan frame. All the results 
have been checked using a direct Jordan frame calculation.

The gauge-Goldstone mode action which we use is
\begin{equation}
{\cal L}_g=-\frac14 F_{a\mu\nu}F^{a\mu\nu}|g|^{1/2}-
\frac12\delta^\perp_{ij}(D_\mu\phi)^i(D^\mu\phi)^j|g|^{1/2}-V(\phi)|g|^{1/2},
\end{equation}
where $D_\mu\phi=\nabla_\mu\phi-gA_{a\mu}T^a\phi$ and $\delta^\perp_{ij}$ is orthogonal
to the background Higgs direction used in the previous section.
 
For the Electroweak theory in particular, variations in the $W$, $Z$ and
the photon directions decouple. In one of the two `$W$' directions for example,
$\varphi^I=( A_{W\mu},\phi^W)$ and the second variation of the action with respect 
to the field variables is
\begin{equation}
-\partial_I\partial_JS_g=
\begin{pmatrix}
-g^{\mu\nu}\nabla^2+\nabla^\mu\nabla^\nu+R^{\mu\nu}+m_W^2&
-m_W\nabla^\mu\\
m_W\nabla^\nu&-\nabla^2+\mu^2+\lambda\phi^2\\
\end{pmatrix}|g|^{1/2},\label{gaugeop}
\end{equation}
where $m_W=g\phi/2$. The second variation in the $Z$ direction is similar, 
with $m_Z=(g^2+g^{\prime 2})^{1/2}\phi/2$, where $g'$ is the coupling to 
the $U(1)$ gauge field. The name `Goldstone modes' has been used, although the 
modes are in fact massive because the background scalar field is not at the 
minimum of the potential.

We can read off the local field-space metric from the coefficients of the Laplacian terms, 
${\cal G}_{IJ}={\rm diag}(g^{\mu\nu},1)|g|^{1/2}$. The
$R_\xi$ gauges from section \ref{sectcov} correspond to the gauge-fixing
functional
\begin{equation}
\chi^\alpha=\nabla^\mu A_{a\mu}-g\phi^TT_a\delta\phi.
\end{equation}
The gauge-fixing contribution to the action is then
\begin{equation}
R_{I\alpha}R_J{}^\alpha=
\begin{pmatrix}
\delta^{ab}\nabla^\mu\nabla^\nu&
m_W\nabla^\mu\\
-m_W\nabla^\nu&m_W^2\\
\end{pmatrix}|g|^{1/2}.
\end{equation} 
The ghost operator in the gauge direction associated with the $W$ is
\begin{equation}
Q^\alpha{}_\beta=-\nabla^2+m_W^2.
\end{equation}
There are two ghosts associated with the $W$ direction and one with the $Z$ direction.

In the covariant approach, there are connection terms in the differential operator 
${\cal D}_I{\cal D}_JS$ because the field-space metric ${\cal G}_{IJ}$ depends on the
spacetime metric, leading to a connection coefficient $\Gamma^K{}_{IJ}$ with $K$ in the
metric direction. The contribution to the operator is 
$E_{\Gamma\,IJ}=\Gamma^K{}_{IJ}\partial_KS$,
\begin{equation}
E_{\Gamma\,IJ}=
\begin{pmatrix}
-G_{\mu\nu}-\kappa^2Vg_{\mu\nu}&0\\
0&\frac12(R-4\kappa^2V)\\
\end{pmatrix}|g|^{1/2},
\end{equation}
where $G_{\mu\nu}+\kappa^2Vg_{\mu\nu}=0$ is the Einstein equation
when $\phi$ is constant. Note that the gauge and the scalar components
of the operator all have a mass term containing $m_W^2+R/2$ when the 
different contributions are added together.

The scaling behaviour of the one-loop action can be found as before by
taking the spacetime background to be the Euclidean four-sphere. 
An important new consideration for the gauge boson beta-functions is the Higgs
field wave function renormalisation at one loop,
\begin{equation}
16\pi^2\gamma_\phi=-\frac34g_{\rm tot}^2+O(\kappa^4V_0)
\end{equation}
where $g_{\rm tot}^2=3g^2+g^{\prime 2}$. The leading term is simply
the flat space result in Landau gauge. The absence of contributions at order
$\kappa^2$ can be seen from relation (\ref{difference}), which expresses the
difference between covariant and non-covariant results in terms of a tensor
polynomial $f$ depending on the operator and the background field Einstein 
equation (\ref{einsteineq}),
\begin{equation}
\mu_R{\partial{\cal L}\over\partial \mu_R}-
\mu_R{\partial{\cal L}_{nc}\over\partial \mu_R}=
f\times\left[\kappa^2(\nabla\phi)^2+4\kappa^2V-R\right]
\end{equation}
The only constant in the operator (\ref{gaugeop}) which could contribute
to $f$ is $\mu^2$. This would give a $\kappa^2\mu^2(\nabla\phi)^2$ term,
but this is of order $\kappa^4V_0$ given our assumptions about $\mu^2$.

\begin{table}[htb]
\def\arraystretch{1.1}
\begin{tabular}{|ll|}
\hline
&Jordan frame\\
\hline
$16\pi^2\beta_\xi$&$-\frac14(6\xi-1) g_{\rm tot}^2+(6\xi-1)\lambda$\\
$16\pi^2\beta_{\mu^2}$&$-\frac12\mu^2g_{\rm tot}^2+6\lambda\mu^2$\\
$16\pi^2(\beta_{\mu^2}+4\kappa^2V_0\beta_\xi)$&
$[6\mu_E^2-4\kappa^2V_0]\lambda-\frac12[\mu_E^2-2\kappa^2 V_0]g_{\rm tot}^2$\\
\hline
\end{tabular}
\begin{tabular}{|ll|}
\hline
&Covariant\\
\hline
$16\pi^2\beta_\xi$&
$-\frac14(6\xi-4)g_{\rm tot}^2+2\lambda$\\
$16\pi^2\beta_{\mu^2}$&
$-\frac12[\mu^2-(8\xi-6)\kappa^2V_0]g_{\rm tot}^2+6[\mu^2+(4\xi-2)\kappa^2V_0]\lambda$\\
$16\pi^2(\beta_{\mu^2}+4\kappa^2V_0\beta_\xi)$
&$[6\mu_E^2-4\kappa^2V_0]\lambda-\frac12[\mu_E^2-2\kappa^2 V_0]g_{\rm tot}^2$\\
\hline
\end{tabular}

\caption{\label{table3} W and Z vector boson contributions to the
$\beta-$functions for the curvature coupling and the mass
of a gravity coupled scalar field at leading order in $\kappa^4V_0$.
The Jordan frame result has been calculated directly from the original action. 
The covariant result uses a geodesic expansion on field space and is independent 
of the frame used. The renormalisation group flow of $\mu_E^2=\mu^2+4\kappa^2V_0\xi$ 
is the same for each of these approaches.
}
\end{table}

\begin{table}[htb]
\def\arraystretch{1.1}
\begin{tabular}{|lll|}
\hline
&Jordan frame&Covariant\\
\hline
$16\pi^2\beta_\lambda$&$6\lambda^2-\lambda g_{\rm tot}^2+\frac38\sum g^4$&
$6\lambda^2-\lambda g_{\rm tot}^2+\frac38\sum g^4$\\
$16\pi^2\beta_6$&
$18\lambda\lambda_6-\frac32\lambda_6g_{\rm tot}^2$&
$18\lambda\lambda_6-\frac32\lambda_6g_{\rm tot}^2+3\xi\lambda g_{\rm tot}^2
-\frac94\lambda g_{\rm tot}^2+9(2\xi-1)\lambda^2$\\
$16\pi^2\gamma_g$&$(6\xi-1)\kappa^2\mu^2$&
$2\kappa^2\mu_E^2-\frac{23}2\kappa^4V_0$\\
\hline
\end{tabular}
\caption{\label{table4}W and Z vector boson contributions to the $\beta-$functions for the 
quartic scalar self-coupling $\lambda$ and the sixth order scalar self-coupling $\lambda_6$
of a gravity coupled scalar field at at leading order in $\kappa^4V_0$. 
The metric wave function anomalous dimension is given to order $\kappa^4V_0$. 
The covariant result uses a geodesic 
expansion on field space and is independent of the frame used.
}
\end{table}

Results are given in tables \ref{table3} and \ref{table4}. The covariant beta-functions
are independent of frame, and differ from the non-covariant expressions. As before,
the two approaches agree on the combination which is important for Higgs stability, 
$\mu^2+4\kappa^2 V_0\xi$. There are differences in the sixth order coupling
$\lambda_6$, but this only enters the Higgs field equations at $O(\kappa^4V_0)$.

We finish of with the top quark as an example of a fermion field. It is far from clear how
the covariant approach generalises to fermion fields, so we take the minimalist approach
and leave off any extra contributions to the effective action. The results are then
checked for consistency against the covariant beta function relations (\ref{betaI}-\ref{betaIII}).
The beta functions are old results, but we repeat them here for completeness. 
The rescaling behaviour
of the effective action due to the quark field is
\begin{equation}
\mu_R{d\Gamma^{(1)}_E\over d\mu_R}=3\,\zeta\left(0,-\nabla^2+m_t^2+\frac14R\right),
\end{equation}
where $m_t=y\phi/\sqrt{2}$ is the top quark mass and $\nabla_\mu$ is the covariant derivative 
acting on Dirac fields. Fermion fields take the positive sign
and the pre-factor takes into account the three colours of the $SU(3)$ gauge group.
With such a simple operator it is best to use general results for the adiabatic expansion
coefficients \cite{Vassilevich:2003xt},
\begin{equation}
b_2(-\nabla^2+E)=2\left(E-\frac16 R\right)^2+O(R^2),
\end{equation}
The wave-function renormalisation at one
loop is $16\pi^2\gamma_\phi=3y^2$. The beta functions inferred from the
renormalisation group equation (\ref{scaling}) are,
\begin{eqnarray}
16\pi^2\beta_\xi\ &=&(6\xi-1)y^2,\\
16\pi^2\beta_{\mu^2}&=&6y^2\mu^2,\\
16\pi^2\beta_\lambda\ &=&-6y^4+12\lambda y^2.
\end{eqnarray}
There is no contribution to $\beta_6$ and $\gamma_g$ at one loop order.

\section{The effective potential and stability}\label{tunnelling}

We have argued that the most important contributions to the corrected Higgs potential are independent of the 
conformal frame and the stability analysis will be unambiguous as long as $\kappa^4V_0$ is negligible. 
To see how this works out in practice, we will consider
instanton induced Higgs vacuum decay during inflation. Higgs instability is caused by 
negative values of the Higgs quartic coupling $\lambda_{\rm eff}$ at large $\phi$. The 
Higgs vacuum is protected by a potential barrier, but the present Higgs vacuum cannot 
survive a period of inflation if quantum fluctuations take the 
Higgs field through the potential barrier. A large, positive, value for $\mu_{\rm eff}^2$  is 
helpful because it reduces the vacuum tunnelling rate \cite{Espinosa:2007qp}.

Perturbative fluctuations in the Higgs field on the scale of the
horizon have a magnitude of order the expansion rate $H$. The vacuum will decay
very rapidly if these fluctuations are larger than the barrier width. If we
denote the value of the field at the maximum of the potential by $\phi_b$, then
rapid vacuum decay occurs for $H>\phi_b$. If $H<\phi_b$, and $|V''(\phi_b)|<4H^2$, 
then the main contribution to vacuum decay can be calculated 
using an instanton solution with the topology of a four-sphere and constant field 
$\phi=\phi_b$ \cite{Hawking:1981fz}.
The instanton induced tunnelling rate $\Gamma_D$ is given by
\begin{equation}
\Gamma_D=Ae^{-B},
\end{equation}
where the dominant effect is due to the exponent, given by the difference in 
action
\begin{equation}
B=S_E[\phi_b]-S_E[0].\label{defB}
\end{equation}
The pre-factor $A$ should be very roughly of order $H^4$ on dimensional grounds,
so that  we can think of $e^{-B}$ as the decay rate per horizon volume per
expansion time.

Consider a classical Higgs action which is given by the Euclidean Jordan frame Lagrangian,
\begin{equation}
{\cal L}_E=\frac12 (\nabla\phi)^2+V(\phi)-{1\over\kappa^2}U(\phi)R.
\end{equation}
The Einstein  equation (\ref{einsteineq}) applied to the instanton four-sphere gives $UR=4\kappa^2V$.
The Lagrangian is constant, and we only need to multiply the Lagrangian by the volume of a four-sphere 
of radius $\sqrt{12/R}$ to get the classical action,
\begin{equation}
S_E[\phi]=-{24\pi^2 U^2\over \kappa^4 V}.
\end{equation}
The same result can be obtained from the Einstein frame action with potential $V_E=V/U^2$.
If we now truncate the difference in action (\ref{defB}) to $O(\kappa^0)$, as in Eqs.
(\ref{vexpand}-\ref{uexpand}), then the contribution to the exponent is
\begin{equation}
B={24\pi^2\over 9H^4}\left[\frac12(\mu^2+12H^2 \xi)\phi_b^2+\frac14\lambda\phi_b^4\right]
\label{Beq}.
\end{equation}
$H$ is the vacuum expansion rate, $3H^2=\kappa^2 V_0$.
The top of the potential barrier is determined by the scalar field equation Eq. (\ref{scalareq}), 
which is equivalent to
\begin{equation}
V_E'(\phi_b)=0,
\end{equation}
Note that all the main features of vacuum decay are determined by the Einstein frame 
potential even when we start out in the Jordan frame. This approach was used
to calculate tunnelling rates in Ref. \cite{Calmet:2017hja}.

We will assume that the quantum corrections are taken into account by
replacing the classical action in the tunnelling exponent with the renormalisation
group corrected effective action,
\begin{equation}
B=\Gamma_E[\phi_b]-\Gamma_E[0],
\end{equation}
where $\Gamma_E$ uses the effective couplings. In particular,
\begin{equation}
V_{\rm eff}=\frac12\mu_{\rm eff}^2(\phi)\phi^2+\frac14\lambda_{\rm eff}(\phi)\phi^4
\end{equation} 
where $\mu_{\rm eff}^2$ is the combination $\mu_{\rm eff}^2=\mu^2+12\xi H^2$.
The couplings are obtained by solving the renormalisation
group equations (\ref{running}).

Initial conditions for the running couplings are set at some chosen point. We take
this point to be $\phi=170{\rm GeV}$, close to the top quark mass.
Combining the results from the tables of beta functions and replacing the vacuum energy by the
expansion rate $H$ gives
\begin{equation}
16\pi^2{d\mu_{\rm eff}^2\over dt}=
12\left(\mu_{\rm eff}^2-2H^2\right)\lambda-\frac12\left(\mu_{\rm eff}^2-6H^2\right)g_{\rm tot}^2
+6\left(\mu_{\rm eff}^2-2H^2\right)y^2,\label{mudot}
\end{equation}
The value of the Higgs coupling is known from experiments at energies less
than $1{\rm TeV}$. The best available values of the Higgs and top quark masses imply
that $\lambda(170{\rm GeV})=0.12577$ \cite{Degrassi:2012ry}.
These experiments are essentially at zero vacuum energy $V_0\approx0$, but
since there is no dependence on the vacuum energy $V_0$ in $\beta_\lambda$, we
can take the experimental values over to the early universe where $V_0$ is large.

The value of $\mu^2$ for the Higgs field as determined in the laboratory is negligible compared to the 
value of $\xi R$ in the inflationary universe, but here we have to take care because of subtleties 
in the properties of light fields in de Sitter space \cite{Weinberg:2005vy}. We already see a hint of 
this in the covariant beta-function which is large, of order $H^2/3$. In the Euclidean approach
to quantum field theory, the infra-red problems lead to a breakdown of perturbation theory
for $\mu_{\rm eff}^2\lesssim\lambda^{1/2}H^2$ \cite{Beneke:2012kn}, so our treatment
will only be valid above this bound. Stochastic approximations imply that the light Higgs field 
develops a mass $\mu_{\rm eff}^2\approx 0.3534\lambda^{1/2}H^2$ 
\cite{Starobinsky:1994bd,Garbrecht:2011gu,Beneke:2012kn,Moss:2016uix}. 
If we assume $\mu^2(170{\rm GeV})\ll H^2$, then this sets a lower limit for de Sitter space 
of $\xi(170{\rm GeV})>0.029\lambda^{1/2}$. We can say nothing about curvature couplings
smaller than this because the techniques required for dealing with
loop corrections with smaller effective mass scales are quite different from the ones we 
use here \cite{Weinberg:2005vy,Kahya:2007bc}.

\begin{figure}[htb]
\begin{center}
\includegraphics[width=0.49\textwidth]{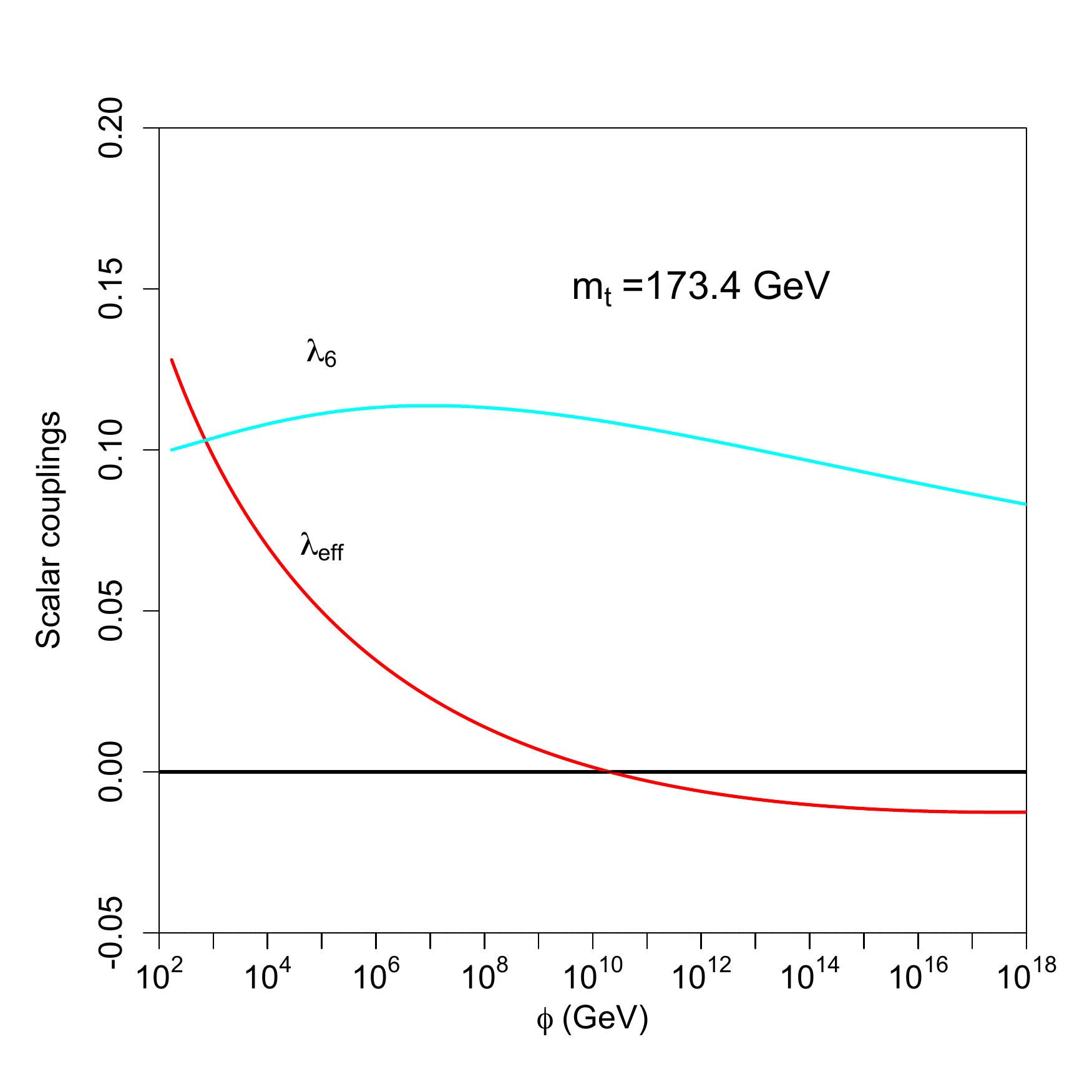}
\includegraphics[width=0.49\textwidth]{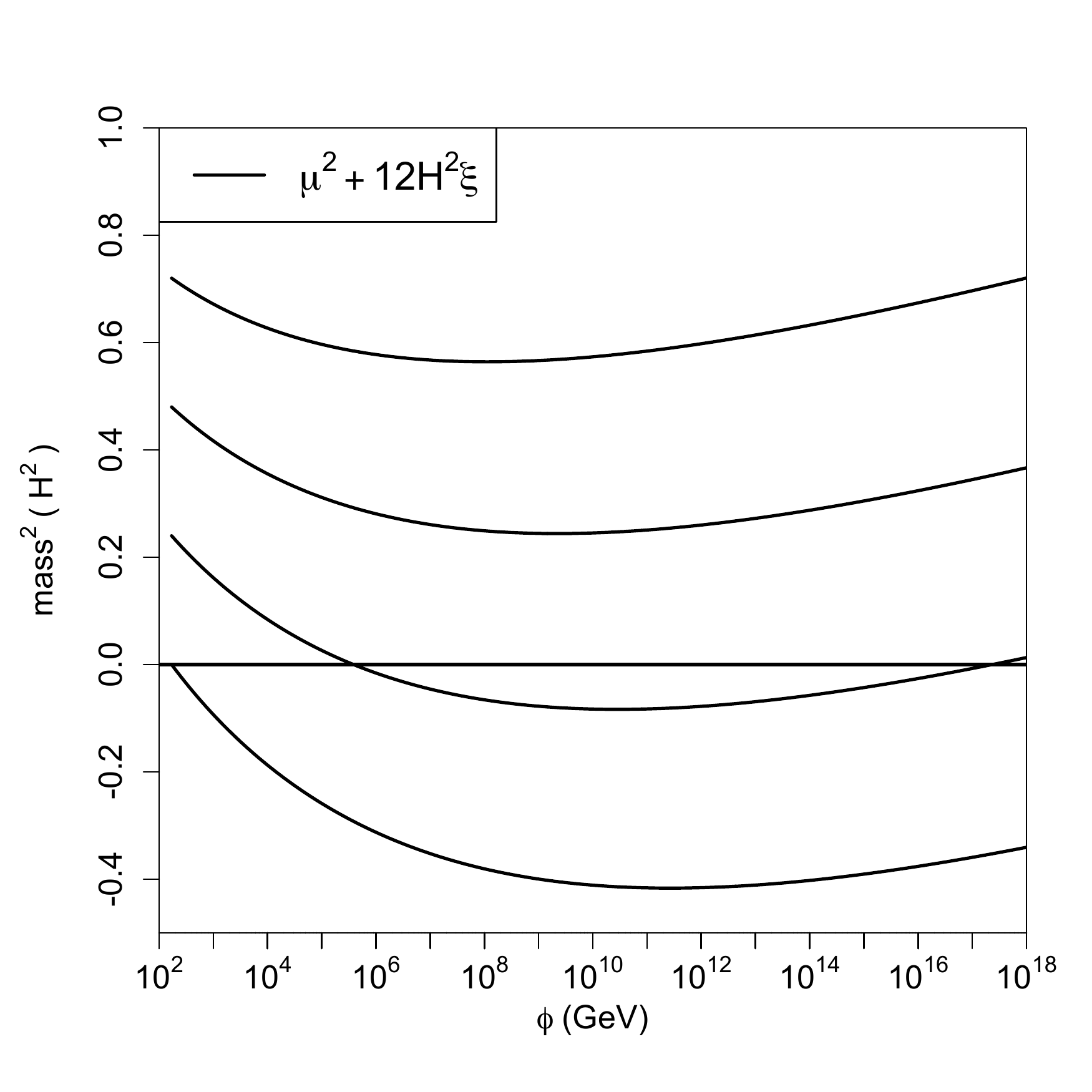}
\caption{
On the left, running couplings $\lambda$ and $\lambda_6$ with $m_t=173.4{\rm GeV}$. 
On the right is the effective mass squared $\mu^2_{\rm eff}=\mu^2+12H^2\xi$. 
The initial conditions are $\lambda(170{\rm GeV})=0.128$,
$\lambda_6(170{\rm GeV})=0.1$ and $\mu^2(170{\rm GeV})=0$.}\label{fig1}
\label{couplings}
\end{center}
\end{figure}

The effective couplings using (\ref{mudot}) are plotted in Fig. \ref{fig1}. 
The standard model couplings $\lambda$, $g$, $g'$ and $y$ have been evolved simultaneously using 
the two-loop flat space beta functions given in \cite{Espinosa:2007qp}. The value of the top 
quark mass $m_t$ sets the scale of the Yukawa coupling $y$ and this has a significant
effect on the running of the Higgs self-coupling, and the value of the field where $\lambda$
vanishes. (The location of the point $\lambda=0$ is not fixed very accurately by the 
renormalisation group corrected potential, and other two loop effects ought to be included.
We have corrected for this by raising $\lambda(170{\rm GeV})$ by 0.2\%.)

\begin{figure}[htb]
\begin{center}
\includegraphics[width=0.5\textwidth]{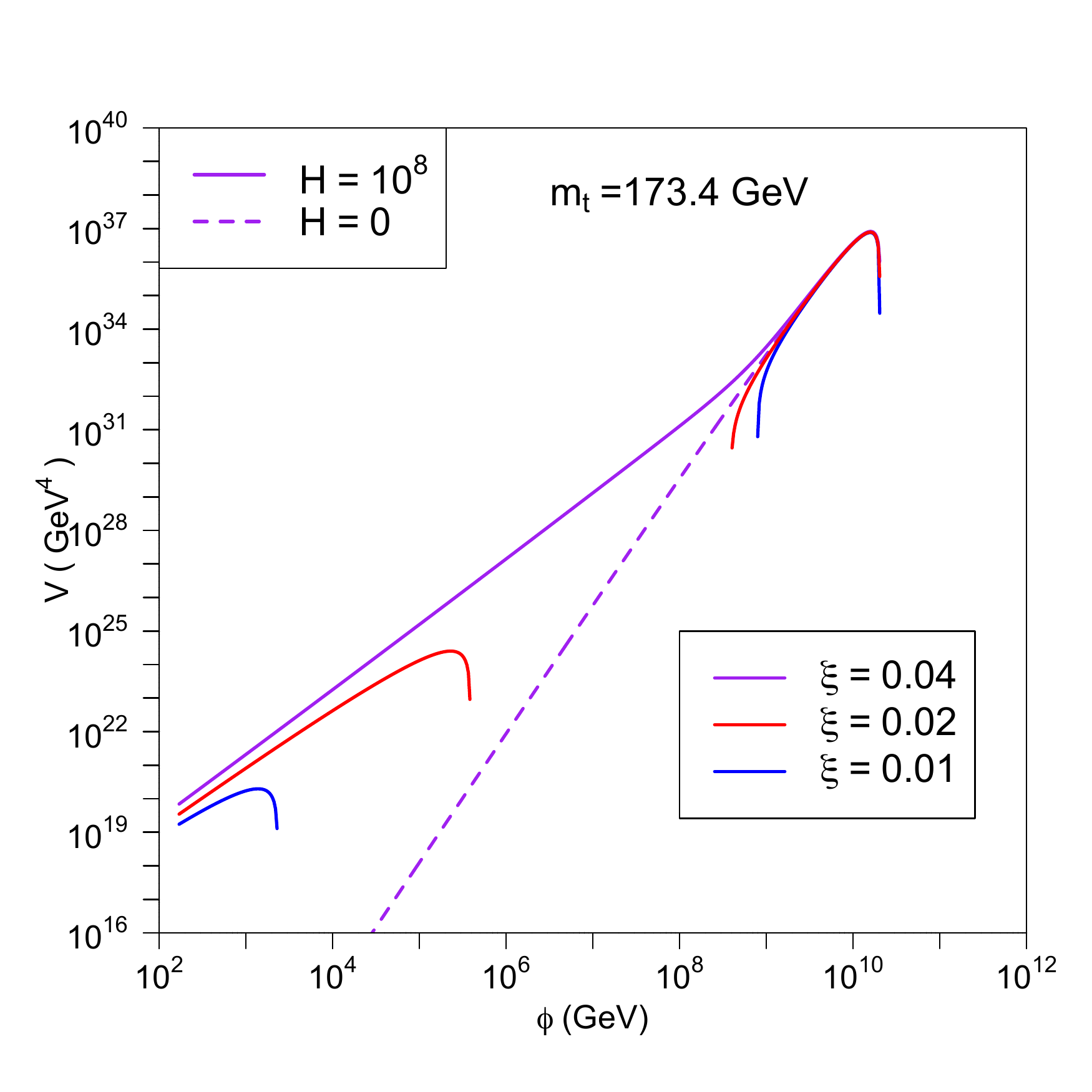}
\caption{The effective Higgs potential plotted as a function of the Higgs field
for $\mu^2=0$ and a range of curvature couplings $\xi$ at $170{\rm GeV}$. 
There is a single maximum for large expansion rate $H$ or $\xi$, but new maxima 
and minima appear for small $H$ and $\xi$.}\label{figP}
\label{couplings}
\end{center}
\end{figure}

Note that for small initial values, the effective mass becomes negative at Higgs
field values below the Planck scale. This can further drive the Higgs instability,
and it can even give the potential a second maximum. This is illustrated by the
potential plots in Fig. \ref{figP}. A combination of small initial $\mu_{\rm eff}^2$
and small expansion rate $H$ leads to twin maxima.

\begin{figure}[htb]
\begin{center}
\includegraphics[width=0.49\textwidth]{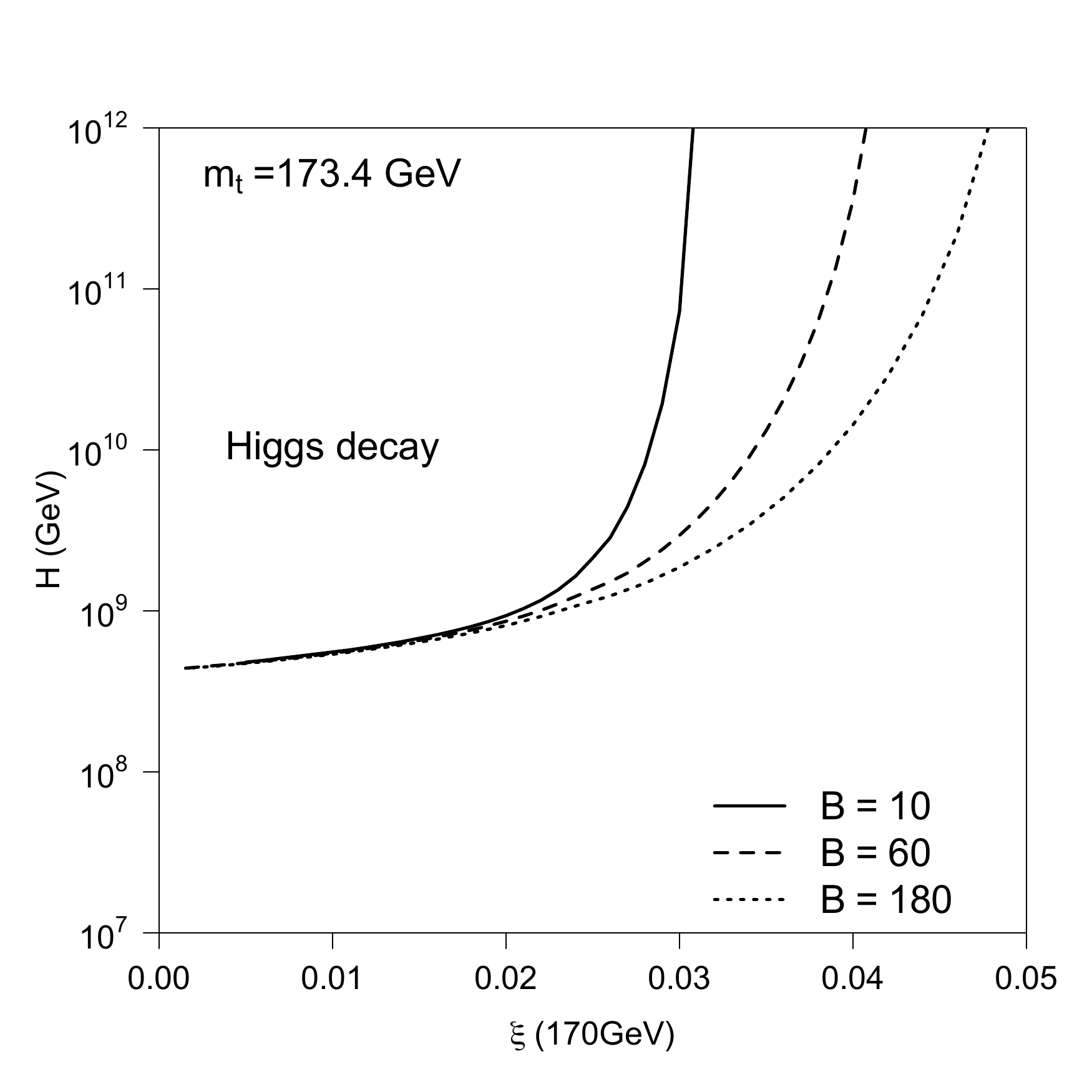}
\includegraphics[width=0.49\textwidth]{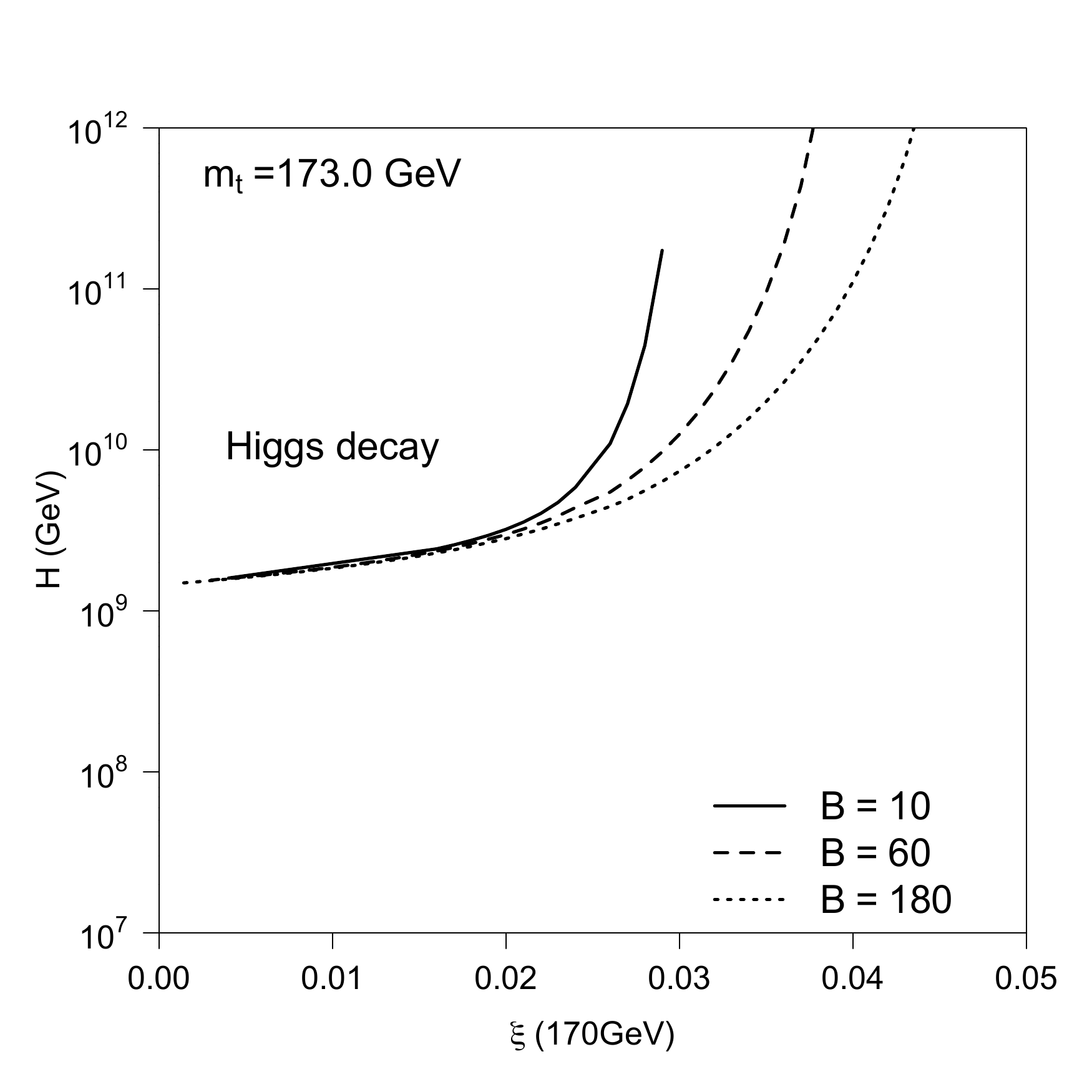}
\caption{Higgs instability regions plotted as a function of the inflationary expansion
rate and the curvature coupling $\xi$ at $170{\rm GeV}$, assuming 
$\mu^2=0$ at $170{\rm GeV}$. Lines show the exponent $B$, related to the vacuum decay 
rate per physical horizon volume, which is of order $e^{-B}$. The Higgs vacuum
is unstable in the region above these curves.
}\label{fig2}
\label{couplings}
\end{center}
\end{figure}

The Higgs effective potential has been used to find the tunnelling exponents 
plotted in Fig. \ref{fig2}. The two plots show different values
of the top quark mass, the first one closest in line with the measured
value and the second which favours Higgs stability. The results are plotted 
against the expansion rate of the Higgs vacuum $H$ and the value of the 
curvature coupling at $170{\rm GeV}$, under the assumption that the Higgs 
mass square $\mu^2$ is negligible at small $\phi$. (This latter assumption is 
questionable, because only the combination $\mu_{\rm eff}^2$ has a frame 
independent meaning when the non-covariant methodology is adopted.)

In the parameter region $\xi<0.022$ the potential has two maxima. Inside this region,
the tunnelling rate is evaluated at the larger maximum. It is sometimes possible for the 
Higgs field to tunnel to the lower maximum, roll down the potential, and then tunnel up 
to the larger maximum. This
combination is less likely than the single tunnelling event, and has no
effect on the stability of the parameter region at the bottom of figure \ref{fig2}.
However, for small curvature coupling, the field could become trapped
between the two maxima during inflation, and return to the present vacuum
state after inflation.

The results support the idea first, proposed in Ref. \cite{Espinosa:2007qp}, that
Higgs stability is sensitive to the value of the curvature coupling, though there are
caveats concerning the case of negative $\xi$, as mentioned above. Running couplings
were included in the Higgs stability analysis of \cite{Herranen:2014cua}. The results are 
consistent with \cite{Herranen:2014cua}, although our results are more precise.
Our main conclusion is that the stability analysis is independent of the
choice of Jordan or Einstein frame.

Exactly how we interpret Fig. \ref{fig2} is dependent on what we understand by the 
inflationary scenario. Suppose first of all that the universe is unique, with the minimum 
amount of inflation of around $N=60$ e-folds. The universe would then have started 
out one de Sitter horizon length across, and have grown to include $e^{3N}$ de Sitter 
horizon regions at the end of inflation. The majority of these had 
to survive Higgs vacuum decay, leading to an extreme limit $B\gtrsim180$. 

In the event that there are many universes like ours, then the conclusion is quite different
because only one out of many universes has to survive Higgs vacuum decay. 
Espinosa et al. \cite{Espinosa:2007qp} 
argued that the survival probability of the universe can be calculated and should be 
$\propto \exp(-H^2 N/32\phi_b^2)$. Assuming that the survival probability of our universe 
should be not be exponentially small gives a lower limit on the field at the top of the potential 
barrier $\phi_b\gtrsim HN^{1/2}$. This condition can be expressed in terms of $B$, for if
we substitute this into Eq. (\ref{Beq}), then
\begin{equation}
B\gtrsim{\pi^2\over 24}{\mu_{\rm eff}^2\over H^2}N.
\end{equation}
In practice $\mu_{\rm eff}^2/H^2\sim 1$, and therefore $\phi_b\gtrsim HN^{1/2}$ corresponds 
very roughly to $B\gtrsim N$.

\section{Conclusion}

The Higgs-gravity system provides an interesting laboratory for trying out ideas
in quantum gravity. One of the issues quantum gravity raises is how to define the
spacetime geometry when scalar and metric backgrounds are allowed to mix freely together.
We have found that applying a methodology which is fully covariant under
such field redefinitions is perfectly feasible. This allows a complete physical equivalence between
the different conformal frames. On the other hand, non-covariant approaches can still be used on scales 
below the Planck mass as long as we are careful. 
We advocate using the effective mass $\mu^2+\xi R$, since this has a 
covariant meaning, and the simpler Einstein frame can always be used.
In particular, we have found that it is always possible to work consistently in a frame
in which the curvature coupling vanishes. The dependence on curvature coupling
in other frames can be recovered from relations like the beta function relations
given in  Sect. \ref{ghtheory}

The effective action calculations we have done allow quite detailed results for Higgs
vacuum decay which take into account running coupling constants, expanding
on the work in \cite{Herranen:2014cua}.
As previously stated \cite{Espinosa:2007qp}, the Higgs-curvature coupling 
can raise the potential barrier around the Higgs vacuum and stabilise the 
Higgs field during inflation. There are some caveats here, such as the extra metastable minima 
in the Higgs potential for small curvature coupling, and infra-red effects which set a 
lower bound to the effective Higgs mass, which deserve further study.

In one respect, the approach adopted has not been as general as it
could, and maybe should, be. The covariant effective action has been
used, but field redefinitions have not been fully integrated with the renormalisation 
group. In a fully general treatment, the renormalised field $\phi_R=Z(\mu_R)\phi$
should become a non-linear mapping into field space, $\phi_R=\phi_R(\phi,\mu_R)$. 
We have have not attempted this, in order to retain as much familiarity with conventional
renormalisation group methods as possible.

Finally, we should point out that we have used existing non-covariant results
for the standard model beta functions which are not associated with the spacetime curvature.
This could cause problems if the running couplings depend on gauge parameters.
In fact, the `$g^2\lambda$' terms in $\beta_\lambda$ are dependent on gauge parameters 
\cite{ParkerTomsbook}. The field value at which the quartic Higgs coupling become negative 
is not protected by any Nielsen identities and may well be gauge parameter dependent. 
We think that may be something to learn from gauge-parameter dependence for Higgs instability in 
flat space. 

\acknowledgments
We would like to acknowledge useful discussions with Gerasimos Rigopoulos.
IGM is supported by the Leverhulme Trust, grant RPG-2016-233 and receives some support 
from the Science and Facilities Council of the United Kingdom, grant number ST/P000371/1.

\appendix

\section{Mode decomposition matrices}\label{matrices}

The differential operators for the gravity-Higgs system reduce to matrices when acting on the sphere
in a tensor harmonic basis. The eigenvalues of the irreducible tensors are $\lambda_T$, $\lambda_V$
and $\lambda_S$ for tensor, vector and scalar harmonics respectively.
The eigenvalues of the matrix $\Delta=\Pi(-\nabla^2+E)\Pi$ and the ghost matrix  $Q$ are used to 
obtain zeta functions and beta functions. These matrices are given explicitly below.

The Laplacian
\begin{equation}
({-\nabla^2})^I{}_J=
\begin{pmatrix}
\lambda_T&0&0&0&0\\
0&\lambda_V-\frac5{12}R&0&0&0\\
0&0&\lambda_S-\frac23R&0&0\\
0&0&0&\lambda_S&0\\
0&0&0&0&\lambda_S\\
\end{pmatrix}.
\end{equation}
The metric on field space,
\begin{equation}
{\cal G}_{IJ}=
\begin{pmatrix}
U&0&0&0&0\\
0&2U(\lambda_V-\frac14R)&0&0&0\\
0&0&\frac34U\lambda_S(\lambda_S-\frac13R)&0&0\\
0&0&0&-4U&-2\kappa U'\\
0&0&0&-2\kappa U'&K+\frac12\kappa^2 U^{\prime 2}/U\\
\end{pmatrix}.
\end{equation}
The non-covariant mass matrix, writing $m^2_T=\frac23 UR-2\kappa^2V$,
\begin{equation}
E_{gIJ}=
\begin{pmatrix}
m^2_T&0&0&0&0\\
0&2m^2_T(\lambda_V-\frac14R)&0&0&0\\
0&0&\frac34m^2_T(\lambda_S-\frac13R)&0&0\\
0&0&0&8\kappa^2V&4\kappa V'\\
0&0&0&4\kappa V'&V''-\frac1{2\kappa^2}RU''\\
\end{pmatrix}.\label{massmat}
\end{equation}
The covariant mass matrix including the connection terms
\begin{equation}
E_{IJ}=
\begin{pmatrix}
m^2_T&0&0&0&0\\
0&2m^2_T(\lambda_V-\frac14R)&0&0&0\\
0&0&\frac34m^2_T(\lambda_S-\frac13R)&0&0\\
0&0&0&8\kappa^2V&2\kappa V'+\kappa^{-1}RU'\\
0&0&0&2\kappa V'+\kappa^{-1}RU'&M^2\\
\end{pmatrix},
\end{equation}
where $M^2=V''-2\kappa^2KV+\frac12KUR-2VU''$.
The gauge transformation matrix
\begin{equation}
{\cal R}^I{}_\alpha=\begin{pmatrix}
0&0\\1&0\\0&2\\0&\frac12\lambda_S\\0&0\\
\end{pmatrix}
\end{equation}.
The projection matrix 
$\Pi^I{}_J=\delta^I{}_J-{\cal R}^I{}_\alpha{\cal N}^{\alpha\beta}{\cal R}_{J\beta}$,
\begin{equation}
\Pi^I{}_J={1\over \lambda_S-\frac12 R}
\begin{pmatrix}
\lambda_S-\frac12 R&0&0&0&0\\
0&0&0&0&0\\
0&0&-\frac12\lambda_S&-2&0\\
0&0&\frac38(\lambda_S-\frac13R)&\frac32(\lambda_S-\frac13R)&0\\
0&0&0&0&\lambda_S-\frac12R\\
\end{pmatrix}.
\end{equation}
The ghost operator
\begin{equation}
Q^\alpha{}_\beta={\cal R}^{I\alpha}{\cal R}_{I\beta}=
\begin{pmatrix}
\lambda_V-\frac14 R&0\\
0&\lambda_S-\frac12 R\\
\end{pmatrix}.
\end{equation}
The ghost metric
\begin{equation}
\gamma_{\alpha\beta}=
\begin{pmatrix}
2U&0\\
0&2U\lambda_S\\
\end{pmatrix}.
\end{equation}

\section{Zeta-function evaluation}\label{az}

Generalised zeta-functions for the operator eigenvalues on a four-sphere can be evaluated by
using a standard binomial expansion method \cite{Dowker:1993ww,Elizalde:1994gf}. 
Eigenvalues of the Laplacian 
are quadratic in $n$, but after diagonalisation of our operators some of the eigenvalues 
are non-polynomial, and a modification of the usual techniques is required.

We start with the Laplacian eigenvalues,
\begin{eqnarray}
\lambda_S&=&\frac{R}{12}\left[\left(n+\frac32\right)^2-\frac94\right]\\
\lambda_V&=&\frac{R}{12}\left[\left(n+\frac52\right)^2-\frac{13}4\right]\\
\lambda_T&=&\frac{R}{12}\left[\left(n+\frac72\right)^2-\frac{17}4\right]\\
\end{eqnarray} 
The degeneracies of these eigenvalues are
\begin{eqnarray}
g_S&=&\frac13\left(n+\frac32\right)^3-\frac1{12}\left(n+\frac32\right)\\
g_V&=&\left(n+\frac52\right)^3-\frac94\left(n+\frac52\right)\\
g_T&=&\frac53\left(n+\frac72\right)^3-\frac{125}{12}\left(n+\frac72\right)\\
\end{eqnarray}
The eigenvalues $\lambda_n$ obtained after diagonalisation are algebraic functions of 
these eigenvalues which can be expanded for large $n$ as power series
\begin{equation}
\lambda_n=\frac1{12}R\left[(n+a)^2+A+B(n+a)^{-2}+\dots\right].
\end{equation}
The degeneracies are generally
\begin{equation}
g_n=b(n+a)^3+c(n+a).
\end{equation}
We replace $\lambda_n^{-s}$ in the zeta-function by its binomial expansion, and
then sums of powers of $n+a$ can be replaced with Hurwitz zeta-functions $\zeta_H(s,a)$,
\begin{equation}
\zeta_H(s,a)=\sum_{n=0}^\infty(n+a)^{-s}.
\end{equation}
The Hurwitz zeta-functions have an analytic extension with a pole at $s=0$ with residue 1. Values at
$s=-1$ and $s=-3$ are Bernoulli polynomials,
\begin{equation}
\zeta(-1,a)=-\frac12B_2(a),\quad
\zeta(-3,a)=-\frac14B_4(a)
\end{equation}

After rearranging the summations we arrive at an expression for the zeta function,
\begin{equation}
\zeta(s)=\left({R\over 12}\right)^{-s}\{bf(s)+cg(s)\},
\end{equation}
where 
\begin{eqnarray}
f(s)&=&\zeta_H(2s-3,a)-sB\zeta_H(2s+1,a)+\frac12s(s+1)A^2\zeta_H(2s+1,a)+\dots\\
g(s)&=&\zeta_H(2s-1,a)-sA\zeta_H(2s+1,a)+\dots
\end{eqnarray}
All of the terms denoted by $\dots$ vanish at  $s=0$ and we are left with
\begin{equation}
\zeta(0)=-\frac12cB_2(a)-\frac14bB_4(a)-{c\over 2}A-{b\over 4}(2B-A^2).\label{zetacont}
\end{equation}

The zeta-function sum is always taken from $n=0$,  but some of the derived modes
are identically zero for some $n$. For example, the gradient of a constant scalar mode
does not give a valid vector mode. These exceptions are handled by subtracting the 
contributions from the $N(h)$ non-existent modes,
\begin{equation}
\zeta(0)=\left(\sum_{n=0}^\infty g_n\lambda_n^{-s}\right)_{s=0}-N(h)
\end{equation}
In particular,
\begin{equation}
N(\nabla_\mu h^S)=1,\quad
N(\nabla_{(\mu} h^V_{\nu)})=10,\quad
N(\nabla_{\mu\nu} h^S)=6,
\end{equation}
the last factor consisting of one $n=0$ and five $n=1$ scalar modes with vanishing
second derivative.

\bibliographystyle{JHEP}
\bibliography{paper.bib}

\end{document}